\documentclass[20pt]{article}
\usepackage{amsmath}
\usepackage{mathtools}
\usepackage{natbib}
\usepackage[margin=0.75in]{geometry}
\usepackage{enumitem}
\usepackage[pdftex]{graphicx}
\usepackage{xcolor}
\begin{document}

\title{Mid-depth Ocean Stratification: Southern Ocean eddies vs interior vertical diffusivity}
\author{Xiaoting Yang and Eli Tziperman}
\maketitle

\section*{Abstract}
The mid-depth ocean stratification was fitted by Munk (1966) to an exponential profile and shown to be consistent with a vertical advective-diffusive balance. However, tracer release experiments show that vertical diffusivity in the mid-depth ocean is an order of magnitude too small to explain the observed 1 km exponential scale. Alternative mechanism suggested that interior diapycnal upwelling is negligible, that all mid-depth water upwells in the Southern Ocean (SO), and that eddies and wind set isopycnal slopes in the SO and therefore determine the mid-depth interior stratification. We examine this hypothesis here by use numerical simulations of eddy-permitting resolution in which we artificially change the diapycnal mixing only away from the SO. We find that SO isopycnal slopes change in response to changes of the interior diapycnal mixing even when the wind forcing is constant. This suggests that SO eddies do not control SO isopycnal slopes via a marginal criticality or a near-vanishing residual overturning conditions but are affected by ocean interior stratification. Furthermore, when the interior mixing is very weak ($10^{-5}$m$^2$s$^{-1}$), the interior stratification is far from exponential, suggesting that SO eddies alone also do not lead to the observed stratification. The results suggest that the communication between the Southern Ocean and the interior is not one-way and that both SO eddies and interior diapycnal mixing are important in determining the interior mid-depth stratification.

\section{introduction}
A balance between upward advection of cold water and downward diffusion of heat was shown by \citet{Munk-1966:abyssal} to be consistent with the exponential-like vertical structure of stratification in the mid-depth Pacific Ocean. In this scenario, some 20 Sv of deep water formation at high latitudes returns to the surface via uniform upwelling, balanced by a vertical diffusivity as large as 10$^{-4}$ m$^2$ s$^{-1}$, leading to the observed one-kilometer exponential depth scale. However, tracer release experiments showed that the actual vertical mixing in the interior mid-depth ocean is an order of magnitude smaller, while much stronger mixing is observed near ocean bottom and boundaries \citep{Ledwell-Watson-Law-1993:evidence, Polzin-Toole-Ledwell-Scmitt-1997:spatial}, due to tidal forcing and breaking of internal waves near topography \citep{Wunsch-Ferrari-2004:vertical}.

This observed boundary enhancement of vertical mixing inspired studies of such mixing on the overturning circulation \citep{Samelson-1998:large, Scott-Marotzke-2002:location}. These studies show that upwelling is concentrated where vertical mixing is enhanced, instead of being uniform over the global ocean. It was therefore suggested by \citet{Miller-Yang-Tziperman-2019:reconciling} that a Munk-like balance in strong-mixing boundary regions, with a large vertical velocity and large diapycnal mixing there, can lead to the observed 1 km scale of the mid-depth (1--3 km) stratification, that is then communicated to the interior via horizontal eddy mixing, as was also suggested by \citet{Munk-Wunsch-1998:abyssal}. While the Munk idea was based on assuming that the vertical mixing coefficient and upwelling are constant in the vertical, \citet{Tziperman-1986:role} showed, and \citet{Miller-Yang-Tziperman-2019:reconciling} verified, that the exponential stratification is, in fact, not sensitive to modest vertical variations in these two variables. In a related research direction, \citet{Ferrari-Mashayek-McDougall-et-al-2016:turning} found that sloping boundaries and enhanced bottom mixing lead to enhanced vertical velocity near the boundaries and possibly to downwelling in the interior. \citet{Baker-Watson-Vallis-2020:meridional} and \citet{Baker-Watson-Vallis-2021:meridional} found that AMOC is responsive to changes of vertical diffusivity as well as SO wind and buoyancy forcings, due to inter-basin exchanges. The exponential fit to the buoyancy profile was found by \citet{Mashayek-Ferrari-Nikurashin-Peltier-2015:influence} to be most accurate in the mid-depth range above the abyssal layer where vertical mixing is bottom-intensified and the vertical variation of $\kappa_v$ is a dominant term in the buoyancy budget.

An alternative approach that has been suggested for explaining the existence of a mid-depth stratification relies on Southern Ocean eddies rather than on vertical mixing. The Southern Ocean is dynamically special in that it is unblocked by meridional continental boundaries above the sill depth. Therefore no geostrophic meridional flows are allowed above the sill because no zonal pressure gradient can be sustained without solid boundaries. This means the northward Ekman transport induced by the westerlies above the Southern Ocean can only be returned southward below the sill depth, geographically or in the bottom boundary layer. This wind-induced overturning is expected to steepen the isopycnal surfaces in the SO. The baroclinic instability associated with these sloping isopycnals generates eddies that act to slump (flatten) the isopycnals. The competition between the wind forcing and eddies in the Southern Ocean therefore results in sloping isopycnals that map the surface density distribution in the SO to a vertical distribution at the northern boundary of the SO, which is the southern boundary of the interior (by ``interior'' we mean north of the SO and away from far north convection sites), setting the interior mid-depth stratification that way \citep{Vallis-2000:large, Wolfe-Cessi-2010:what, Nikurashin-Vallis-2011:theory}. In the Munk picture the mid-depth stratification vanishes at the limit of vanishing vertical mixing. When these SO eddy dynamics are taken into account, the mid-depth stratification still does not vanish, even if the overturning circulation is considered adiabatic and the interior diabatic diffusivity is neglected. In this adiabatic limit, the MOC return flow to the surface occurs along sloping isopycnals in the SO \citep{Vallis-2000:large, Wolfe-Cessi-2010:what, Nikurashin-Vallis-2011:theory}. However, it has been pointed out by \citet{Miller-Yang-Tziperman-2019:reconciling} that the mid-depth stratification in this case is not exponential. \citet{Nikurashin-Vallis-2011:theory} pointed out that in the presence of large vertical diffusion, the stratification is determined by the Munk balance rather than by the SO wind and eddies. They do not discuss the issue of exponential stratification, and suggest that this regime may be relevant to the abyssal stratification rather than mid-depth one.

The idea that eddies strongly influence isopycnal slopes is of critical importance for the above studies showing how eddies can lead to a mid-depth interior stratification in the adiabatic overturning picture, and two mechanisms were proposed to explain the relation between eddies and isopycnal slopes more generally. One involves the idea of marginal criticality or eddy saturation, and the other of a fully compensated residual circulation, and both are discussed below. In the idealized limit in which the SO eddies are strong and the interior diapycnal mixing is weak, both of these ideas lead to an isopycnal slope in the SO that depends only on the SO winds and eddy dynamics. This means that the mid-depth ocean stratification north of the SO is then fully determined by the SO wind and eddies. These idealized limits are useful extreme states which we examine in this paper in order to identify their relation to the observed interior exponential mid-depth stratification.

The first idea is that slopes are at the so-called marginal criticality \citep{Stone-1978:baroclinic, Jansen-Ferrari-2012:macroturbulent, Jansen-Ferrari-2013:equilibration, Jansen-Ferrari-2013:vertical}. At marginal criticality, larger slopes would lead to eddy formation via baroclinic instability and the eddies would then draw APE from the mean slopes, and therefore flatten them back to marginal criticality. This marginal criticality may be diagnosed via the ratio between the Rhines scale and the deformation radius \citep{Jansen-Ferrari-2012:macroturbulent}, or via the ratio between the PV gradient due to layer thickness variations and $\beta$ effect \citep{Jansen-Ferrari-2013:equilibration}. If eddies are sufficiently efficient in drawing APE from the mean slope, to maintain a state whose criticality is approximately one, isopycnal slopes are completely determined by the competition between the process that tend to steepen isopycnals (e.g., SO winds) and the eddies that slump them. However, it has been suggested that in the ocean eddies are not expected to be sufficiently effective in this sense, and slopes are therefore not necessarily at marginal criticality \citep{Jansen-Ferrari-2012:macroturbulent}. A related concept to this criticality is that of eddy saturation, where as the SO winds get stronger the isopycnal slopes stop responding due to the effect of ocean eddies that keep the slopes at saturation \citep{Munday-Johnson-Marshall-2013:eddy}.

The second idea regarding the role of eddies in affecting SO isopycnal slopes, is that eddies set these slopes via buoyancy advection in opposite direction to the wind-driven overturning. Specifically, it has been shown that if the SO residual circulation (nearly) vanishes, isopycnal slopes are determined by a balance between SO wind forcing and eddies \citep{Gnanadesikan-1999:simple, Marshall-Radko-2003:residual, Wolfe-Cessi-2010:what, Nikurashin-Vallis-2011:theory}. It has also been shown that the residual circulation is expected to vanish in the limit of zero interior diffusivity and in the absence of surface buoyancy forcing in the SO \citep{Marshall-Radko-2003:residual}. In this limit, SO eddies and winds would be setting the SO isopycnal slopes and therefore the interior mid-depth ocean stratification without contribution by other physical processes away from the SO. However, \citet{Wolfe-Cessi-2010:what} showed in eddy-resolving numerical simulations that as the interior diffusivity vanishes, there is a strong reduction of the SO residual circulation, to about 20\% of the Eulerian overturning circulation, so that the residual circulation does not completely vanish. Furthermore, the residual overturning and eddy activity in the SO \citep[and therefore the equivalent GM coefficient,][]{Gent-Mcwilliams-1990:isopycnal} have been found to be sensitive to changes in vertical mixing in idealized GCM simulations \citep{Munday-Johnson-Marshall-2013:eddy}.

We can summarize the role of the residual circulation identified in previous studies as follows. \citet{Wolfe-Cessi-2010:what} find that the dynamics of the SO is nearly adiabatic beneath the surface layer, and that its isopycnal slopes, and thus the middepth interior ocean stratification as well, does not depend strongly on the value of the interior diffusivity. Similarly, \citet{Nikurashin-Vallis-2011:theory} find that for weak diapycnal interior mixing typical of the middepth ocean, the middepth overturning circulation and stratification are primarily controlled by the wind and eddies in the SO channel, with diapycnal mixing playing a minor role. In the limit of weak diapycnal interior mixing, the interior stratification does not vanish and is determined by the SO winds and eddies (as represented by the GM coefficient).

In this paper, we accept the above premise that the SO eddies allow for a non-vanishing mid-depth stratification in the limit of weak interior diapycnal mixing. We then show, though, that for this stratification to be exponential, as robustly observed \citep{Munk-1966:abyssal, Miller-Yang-Tziperman-2019:reconciling}, diapycnal ocean mixing is necessary and should not be ignored. We use numerical simulations of eddy-permitting resolution to show that the Southern Ocean isopycnal slopes and eddies are influenced by the interior vertical diffusivity and that the eddies cannot determine the SO slopes regardless of processes in the ocean interior north of the SO. This means that processes in the interior are still important in determining the structure of the mid-depth stratification. For this purpose, we change values of the diapycnal diffusivity, $\kappa_v$, only north of the SO channel, in a range of small to unrealistically large values, to study the possible response of the SO isopycnal slopes to changes in the interior basin diapycnal mixing (section ~\ref{sec:res}\ref{sec:SO}). We find that both isopycnal slopes and eddy activity in the SO, as reflected in measures of both the marginal criticality and residual circulation, are responsive to changes in the interior mixing. We also show, similar to the coarser, non eddy-resolving model of \cite{Miller-Yang-Tziperman-2019:reconciling}, that when the interior vertical diffusivity is small, the resulting adiabatic MOC dynamics and SO eddies are unable to lead to an observed-like interior mid-depth exponential vertical stratification. We conclude that the exponential mid-depth interior ocean stratification must be determined by an interplay of interior vertical diffusion and SO eddy dynamics.

%XX To be completed later: an accompanying paper submitted in parallel to this one discusses\ldots

This paper is organized as follows. We describe the experimental designs in section \ref{sec:model}. Results are presented in section \ref{sec:res} which is further divided into a discussion of the mid-depth stratification (section~\ref{sec:res}\ref{sec:strat}), and the response of the Southern Ocean  to changes of interior vertical mixing (section~\ref{sec:res}\ref{sec:SO}). And we conclude in section~\ref{sec:conclusion}.

\section{Model configurations}
\label{sec:model}

\begin{figure}[t]
  \centering
  \noindent\includegraphics[width=0.75\paperwidth]{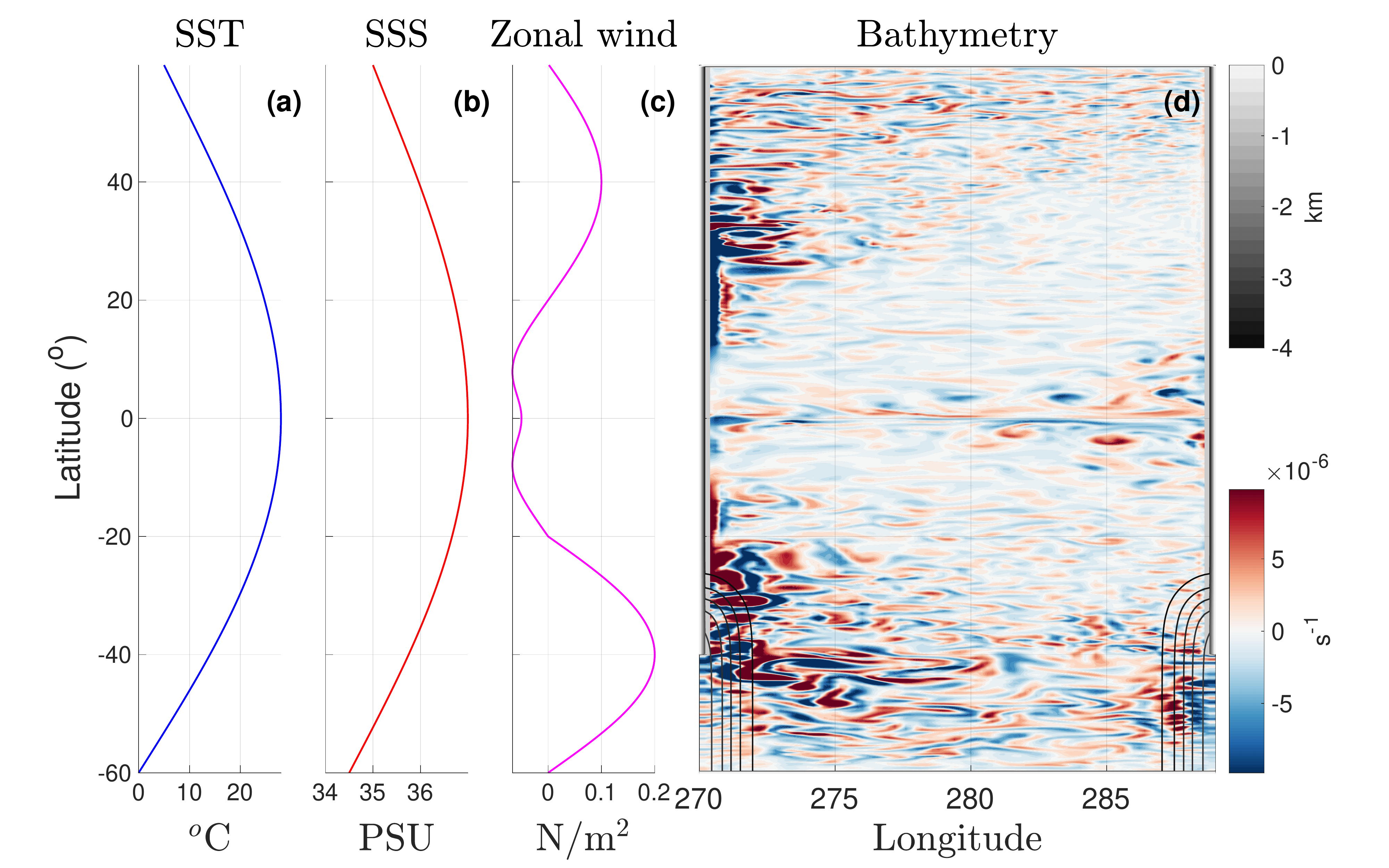}
  \caption{The model configuration. (a) SST restoring profile; (b) SSS restoring profile; (c) zonal wind stress; (d) gray-scale contours: bathymetry (km); colors: a snapshot of surface relative vorticity in the default case.}
  \label{model_setup}
\end{figure}

We use an idealized configuration of Massachusetts Institute of Technology General Circulation model \citep[MITgcm, ][]{Marshall-Adcroft-Hill-et-al-1997:hydrostatic}. The idealized geometry and bathymetry of our simulations are shown in Fig.~\ref{model_setup}d. This domain ranges from 60$^\circ$S to 60$^\circ$N, with an re-entrant channel between 60$^\circ$S and 40$^\circ$S. In the channel, there is a smoothed sill at the depth of 3 km. This sill is used to guarantee that the transport of the circumpolar current is within a realistic range ($\sim$100 Sv), and it is smoothed to avoid spurious numerical noise in the vertical velocity that may influence stratification, which is the focus of this paper. Otherwise the bathymetry is flat at the depth of 4.2 km. The horizontal resolution is 0.2$^\circ\times$0.2$^\circ$, and we use 51 vertical layers in total, whose thickness ranges from 5 meters near the surface to 125 meters near the bottom. Due to the relatively high horizontal resolution, we use a narrow basin that is only 20 degree wide to limit the computational cost.

The model is forced at the surface by restoring to longitude-independent temperature and salinity fields using a restoring time scale of 10 days, and with a zonal wind stress (Figs.~\ref{model_setup}a,b,c). The parameters that are constant for all cases include horizontal harmonic viscosity ($A_h=10$ m$^2$ s$^{-1}$), horizontal bi-harmonic viscosity ($A_4=1.5\times 10^{10}$ m$^4$ s$^{-1}$) , vertical viscosity ($A_v=10^{-4}$ m$^2$ s$^{-1}$), horizontal bi-harmonic diffusivity ($\kappa_4=1.5\times 10^{10}$ m$^4$ s$^{-1}$) and a linear bottom drag coefficient ($r_b=2\times 10^{-4}$ s$^{-1}$). These parameters are summarized in Table~\ref{model_param}. We use a nonlinear equation of state following \citet{Jackett-Mcdougall-1995:minimal}. Considering the high-resolution used here, therefore eddies are resolved (a snapshot is shown in Fig.~S1), and the \citet{Gent-Mcwilliams-1990:isopycnal} scheme is therefore not used. Convection is represented using the K-profile parameterization scheme \citep{Large-Mcwilliams-Doney-1994:oeceanic}, and we use the 3rd order DST Flux Limiter for advection (33 in MITgcm).

We focus on the effect of vertical mixing in the domain away from the SO, north of 40$^\circ$S, on stratification and SO eddies. Therefore, the vertical diffusivity parameter is unchanged from its base value (10$^{-5}$ m$^2$ s$^{-1}$) in the Southern Ocean for all cases. We use a series of values for this vertical diffusivity north of the channel: 10$^{-5}$, 10$^{-4}$, $2\times 10^{-4}$, $4\times 10^{-4}$, $6\times 10^{-4}$, $8\times 10^{-4}$ and $10^{-3}$ m$^2$ s$^{-1}$. A 5-degree wide latitudinal range north of 40$^\circ$S linearly links the value of $\kappa_v$ in the channel and the interior (Table~\ref{case_series}). This design to only change vertical diffusivity north of the channel is artificial because $\kappa_v$ is a measurement of how strong the interactions of tidal forcing and internal waves with solid boundaries are in the ocean  \citep{Wunsch-Ferrari-2004:vertical} and therefore should be increased systematically throughout the domain. We also explore a range of $\kappa_v$ that is unrealistically large. However, this set-up is consistent with the purpose of this paper to study the possible response of Southern Ocean isopycnal slopes and eddies to changes in interior stratification.

We run the simulations until a statistical steady state is reached (time series of layerwise-averaged temperature and salinity converge) and then run the model for 20 more years. The temporally averaged model variables ($T$, $S$, velocities) are calculated using model output from these last 20 years. Nonlinear terms in the temperature budget and in the Lorentz energy cycle are averaged using bi-daily data over the last ten years.

\section{Results}
\label{sec:res}

This section is divided into two parts: the response of mid-depth stratification to changes in interior vertical diffusivity is discussed in section~\ref{sec:res}\ref{sec:strat}, and the response of Southern Ocean eddies in section~\ref{sec:res}\ref{sec:SO}. In section~\ref{sec:res}\ref{sec:strat} we show that the quality of an exponential fit to the vertical profile of stratification (vertical derivative of potential density) improves with an increasing interior vertical diffusivity $\kappa_v$. At the same time, isopycnal slopes in the SO increase and SO eddy activity strengthens as the interior $\kappa_v$ increases. We explain that this suggests that SO eddies are not able to maintain the SO stratification at marginal criticality, which we later use to discuss the processes responsible for the mid-depth interior ocean stratification. We then explain the eddy response by analyzing the Eady growth rate and the Lorenz energy cycle in the Southern Ocean (section~\ref{sec:res}\ref{sec:SO}).

\subsection{Mid-depth stratification}
\label{sec:strat}

\begin{figure}[t]
  \centering
  \noindent\includegraphics[width=0.75\paperwidth]{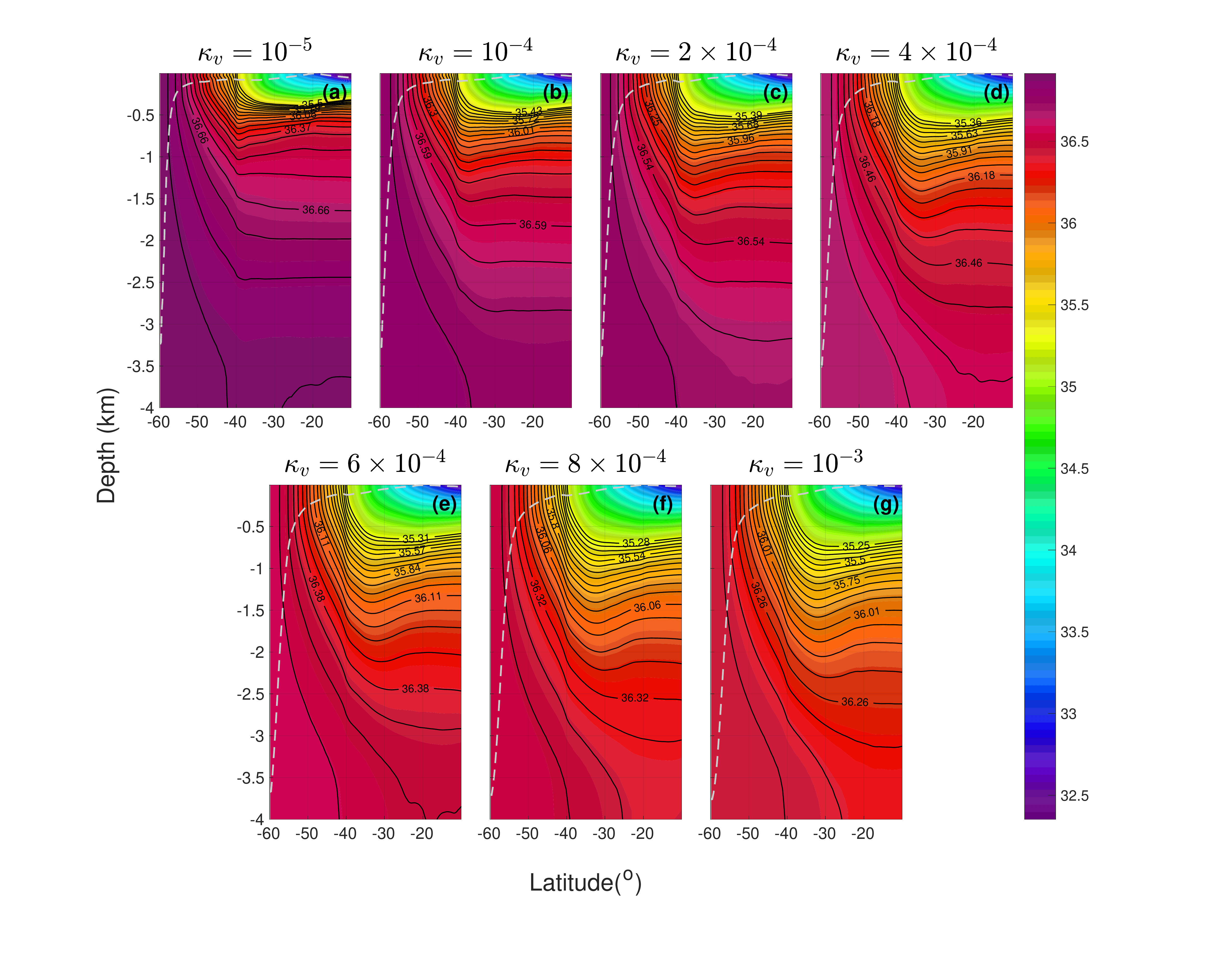}
  \caption{Colors: zonally averaged $\sigma_2$ south of 15$^\circ$S. Black contours: 20 isopycnal surfaces that originate from the surface of the southern channel. Gray dashed line: bottom of the surface mixed layer, output from the K-profile parameterization package. (a) control case; (b) Kvx10; (c) Kvx20; (d) Kvx40; (e) Kvx60; (f) Kvx80; (g) Kvx100.}
  \label{south_strat}
\end{figure}

\begin{figure}[t]
  \centering
  \noindent\includegraphics[width=0.75\paperwidth]{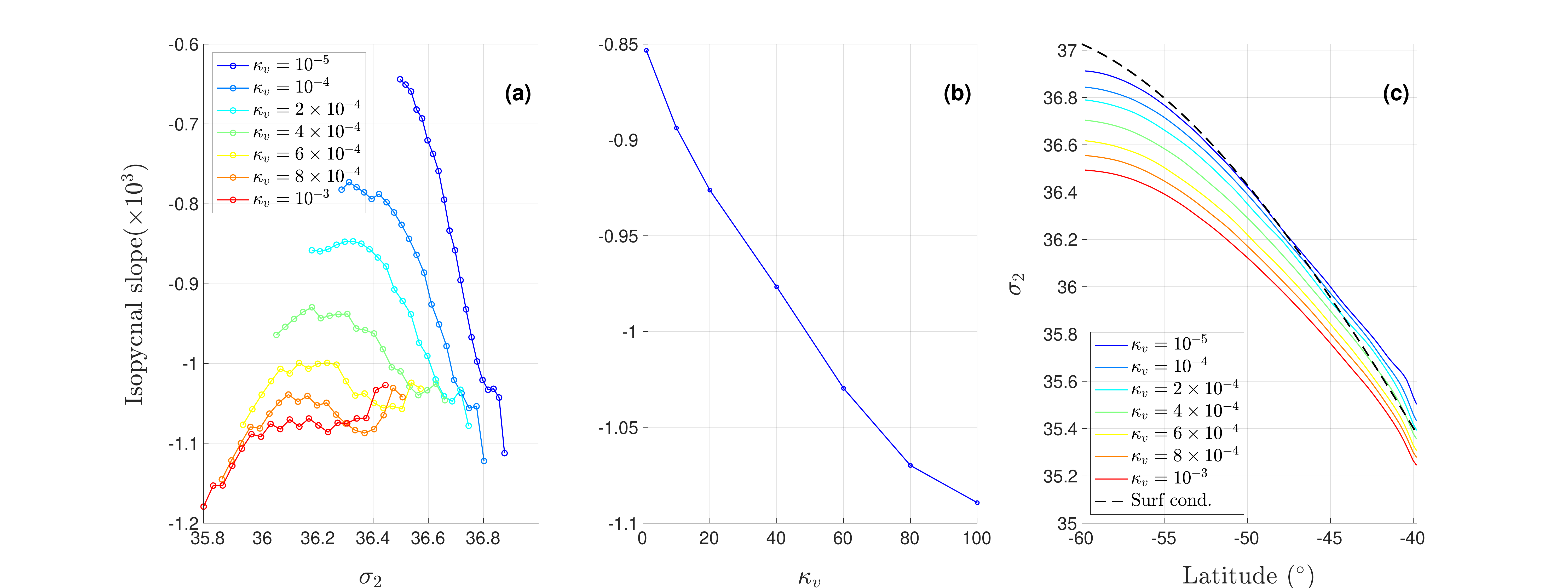}
  \caption{The averaged isopycnal slopes in the channel, calculated using the 20 isopycnal surfaces that outcrop in the SO channel and exist between 1 end 3.5 km depth at 40$^\circ$S. (a) Slopes for the twenty isopycnal surfaces, in all cases (color); (b) Averaged isopycnal slope in the channel as a function of interior $\kappa_v$; (c) Zonally averaged $\sigma_2$ at the surface in the channel for all cases (colored lines) and the surface restoring boundary condition (dashed black line).}
  \label{slope_line_plot}
\end{figure}

As the vertical diffusivity in the interior is increased, the mid-depth ocean gets warmer and therefore lighter as more heat diffuses downward (Figs.~\ref{south_strat}, S2). In Fig.~\ref{south_strat}, twenty isopycnals that both outcrop in the SO channel and exist in a vertical density distribution at 40$^\circ$S are plotted (solid black lines). It is clear that the slopes of these isopycnal surfaces in the SO become steeper as the interior $\kappa_v$ is increased. With stronger downward heat diffusion, the interior warms and isopycnals corresponding to larger density move downward. For isopycnals to be able to link the interior density distribution to the prescribed surface density distribution at the surface of the channel, these SO isopycnals must be steeper in the higher interior diffusivity runs. At the same time, the SO surface also becomes lighter (Fig.~\ref{slope_line_plot}c). This will partly counteract the steepening effect discussed above. The net effect of these two mechanisms is steeper isopycnals surfaces in the southern channel (Fig.~\ref{slope_line_plot}a,b). Fig.~\ref{slope_line_plot}a clearly shows that the slopes of the isopycnal surfaces that connect the surface of the southern channel and the southern boundary of the interior basin between 1 and 3.5 km depth are systematically steeper as interior $\kappa_v$ is increased.  Furthermore, Fig.~\ref{south_strat}a shows that when interior $\kappa_v$ is small, the isopycnal slopes become steeper for denser water masses, consistent with stronger deep water formation near the southern solid boundary (Fig.~\ref{south_strat}). When interior $\kappa_v$ is large (in the ``Kvx80'', and ``Kvx100'' cases), the slopes for different isopycnal surfaces are comparable.

The criticality parameter in the atmosphere or the Southern Ocean can be defined as $\xi\equiv fs/\beta H$, where $f=2\Omega\sin\theta$ is the Coriolis parameter, $\beta$ is the meridional gradient of $f$, $s$ is the isentropic or isopycnal slope and $H$ is the layer thickness. In the atmosphere, the isentropics that intersect with the surface in the tropics reaches the top of the troposphere near the poles, implying a state of marginal criticality \citep[$\xi\approx 1$,][]{Jansen-Ferrari-2013:equilibration}. Such conclusions cannot be drawn for the Southern Ocean. The observed response of the isopycnal slopes in the Southern Ocean to interior $\kappa_v$ changes implies that SO eddies are not sufficient to maintain a state of marginal criticality, consistent with \citet{Jansen-Ferrari-2012:macroturbulent}. In fact, we found that SO in our simulations is in a supercritical state and criticality increases as interior $\kappa_v$ increases (Appendix B, Fig.~S6).

\begin{figure}[t]
  \centering
  \noindent\includegraphics[width=0.75\paperwidth]{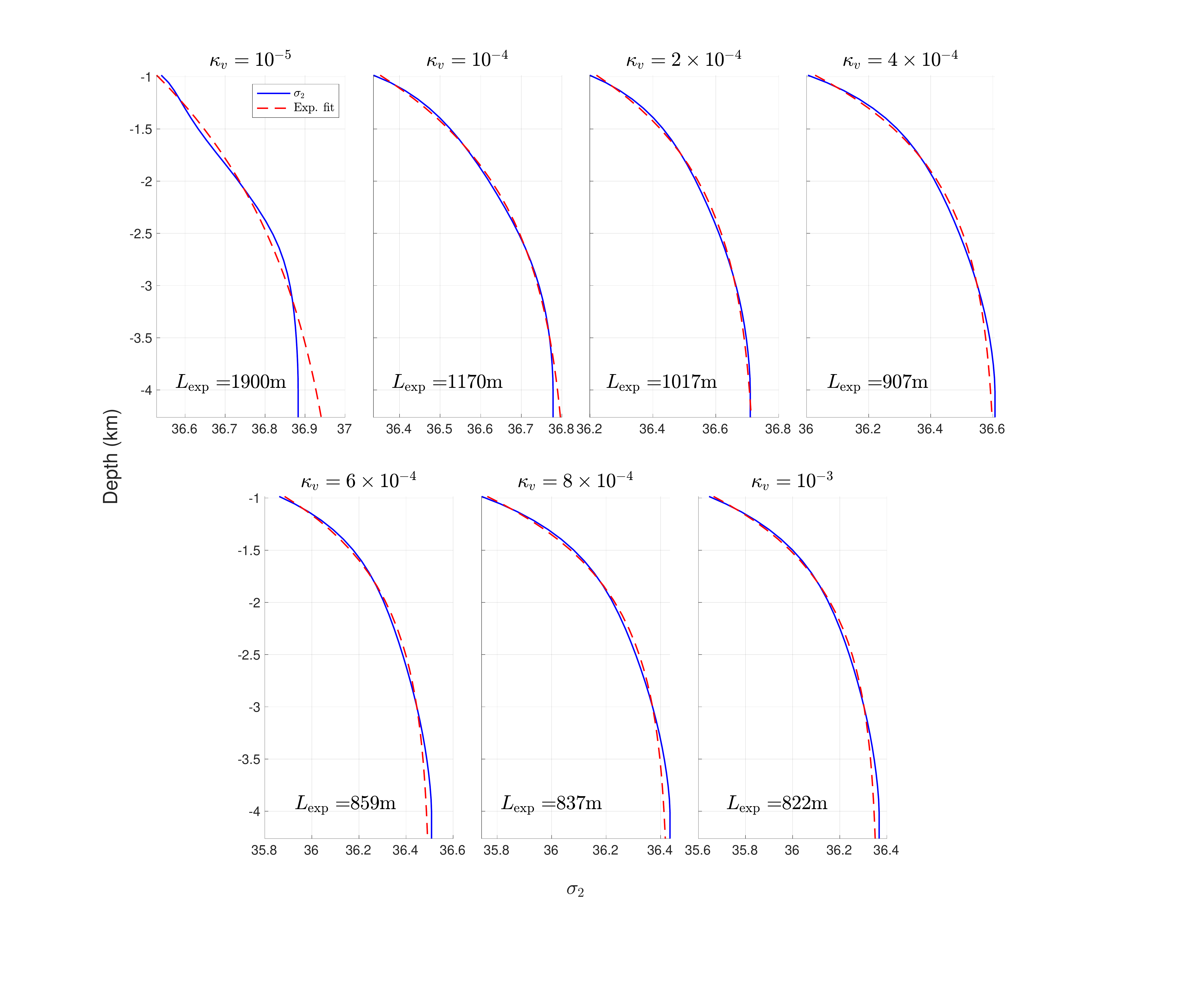}
  \caption{Exponential fit of the vertical profiles of $\sigma_2$ averaged between 30$^\circ$ S to 30$^\circ$ N, three degrees from the western and the eastern boundaries. The fit coefficients are calculated only between 1-3 km depth, and the figure is extended to the bottom. Blue solid line: model output; Red dashed line: exponential fit. $L_{\mathrm{exp}}$: exponential decay length scale. (a) control case; (b) Kvx10; (c) Kvx20; (d) Kvx40; (e) Kvx60; (f) Kvx80; (g) Kvx100.}
  \label{rho_exp_fit}
\end{figure}

The interior vertical profiles of different variables in this section is calculated between 30$^\circ$S and 30$^\circ$N, three degrees from the western and the eastern boundaries. Fig.~\ref{rho_exp_fit} show the vertical profiles of $\sigma_2$ output by the model (solid blue lines) and the corresponding exponential fit (dashed red lines) calculated between one and three kilometers depth but plotted all the way to the bottom of the domain. The errors in the exponential fit of the profiles of $\sigma_2$ are small even in the limit of small interior $\kappa_v$ (default case). At the same time, the exponential decay scale ($L_{\mathrm{exp}}$) is decreasing as stronger vertical mixing is applied, consistent with \citet{Munk-1966:abyssal}. It is also noted that $L_{\mathrm{exp}}$ is in a range that agrees well with the observed one kilometer scale in simulations where $\kappa_v$ does not exceed $4\times 10^{-4}$m$^2$ s$^{-1}$.

\begin{figure}[t]
  \centering
  \noindent\includegraphics[width=0.75\paperwidth]{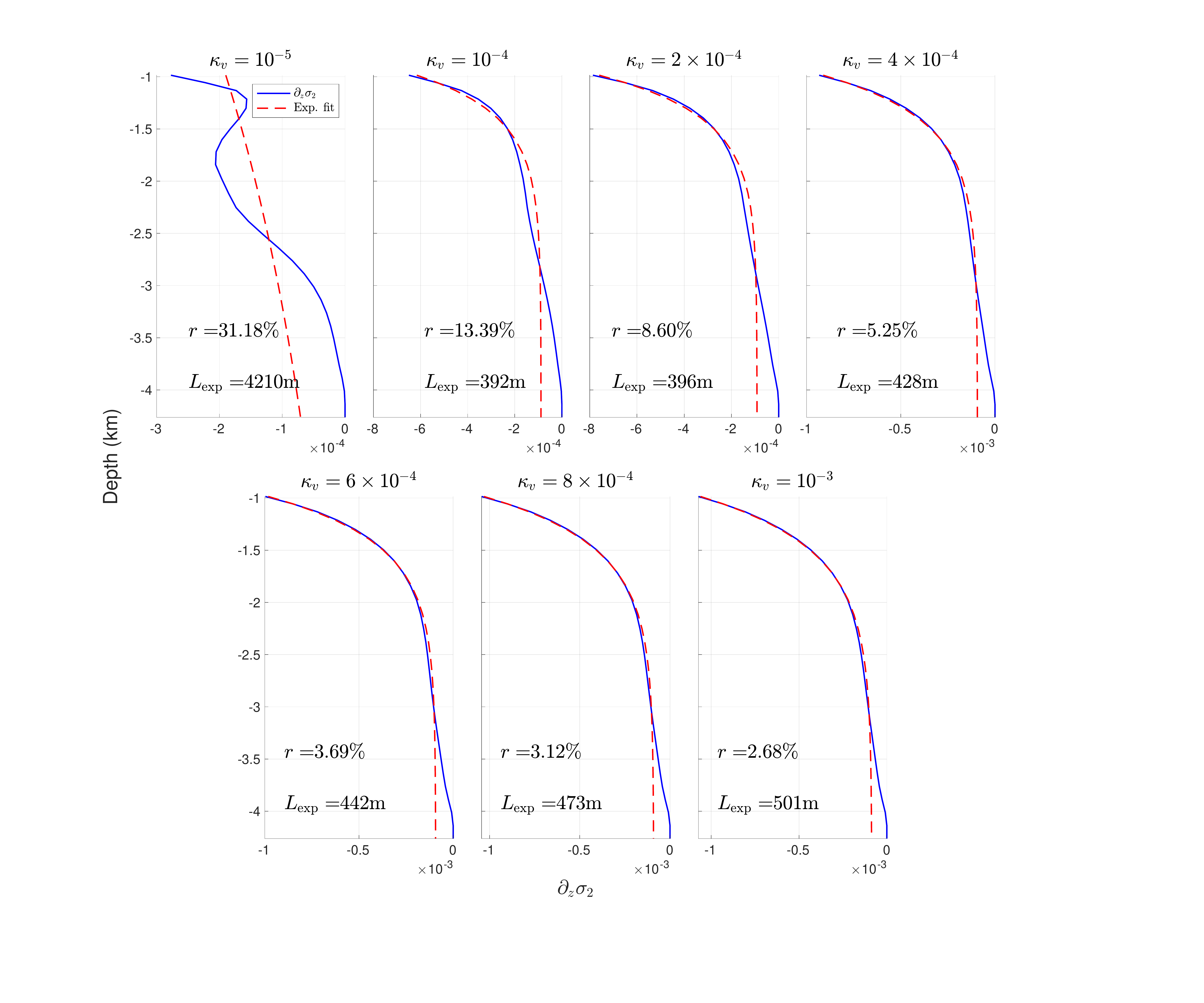}
  \caption{Exponential fit of the vertical profiles of $\partial_z\sigma_2$ averaged between 30$^\circ$ S to 30$^\circ$ N, three degrees from the western and the eastern boundaries. The fit coefficients are calculated only between 1-3 km depth, and the figure is extended to the bottom. Blue solid line: model output; Red dashed line: exponential fit. $L_{\mathrm{exp}}$: exponential decay length scale. $r$: the relative error between the model output and exponential fir, $r=\sqrt{\sum\left( \rho_z(k)-\rho_{z,\mathrm{exp}}(k) \right)^2 dZ(k)/\sum \rho_z(k)^2 dZ(k)}$. (a) control case; (b) Kvx10; (c) Kvx20; (d) Kvx40; (e) Kvx60; (f) Kvx80; (g) Kvx100.}
  \label{rho_z_exp_fit}
\end{figure}

However, the quality of the exponential fit for stratification profiles ($\partial_z \sigma_2$) is poor at low vertical mixing values and improves considerably at stronger vertical mixing in the $\kappa_v$ range used in these simulations (Fig.~\ref{rho_z_exp_fit}). The fitted profile of $\partial_z \sigma_2$ is almost a constant (with a exponential length scale of 4210 m for a fit between 1--3 km depth range) in the default simulation. And the stratification profile itself indeed does not show an exponential shape in this case but rather is wiggly. To quantify how well the exponential fits agree with the model output, we define the relative error between these two profiles as
\begin{equation}\label{relative_err}
  r=\sqrt{\frac{\sum_k\left( \rho_z(k)-\rho_{z,\mathrm{exp}}(k) \right)^2 dZ(k)}
    {\sum_k \rho_z(k)^2 dZ(k)}},
\end{equation}
in which $k$ is the vertical model level index, $\rho_z$ is the stratification calculated from the model output, $\rho_{z,\mathrm{exp}}$ is that from the exponential fit, and $dZ$ is the thickness of the $k$th layer. The $r$ values are written in Fig.~\ref{rho_z_exp_fit}. We observe a steady improvement of the agreement between the model output profile and the exponential fit, as interior $\kappa_v$ is increased.

\citet{Munk-1966:abyssal} assumed a temperature budget between upward mean flow advection and downward diffusion that works when $\kappa_v\sim 10^{-4}$m$^2$ s$^{-1}$, while \citet{Wolfe-Cessi-2010:what} found a dominant role played by eddy transport in the limit of small vertical diffusivity. The temperature budget (eq.~\ref{temp_budget}) at steady state is analyzed and the vertical profiles of each term is shown in Fig.~S3,
\begin{equation}\label{temp_budget}
\overline{\mathbf{u}_H}\cdot\nabla_H\overline{T}+\overline{w}\partial_z\overline{T}+\overline{\mathbf{v^\prime}\cdot\nabla T^\prime}=\text{Horiz diff}+\text{Vert diff}.
\end{equation}
In this equation, $\overline{(\cdot)}$ stands for a time average and $(\cdot)^\prime$ is a deviation from time average. The symbol $\mathbf{u}_H$ denotes the horizontal velocity vector, $\mathbf{v}$ is the 3d flow field vector, $w$ is vertical velocity and $T$ is the temperature. $\nabla_H$ and $\nabla$ are the horizontal and 3d gradient operators, respectively. And ``Horiz diff'' and ``Vert diff'' terms are model horizontal and vertical diffusion terms due to parameterized eddy diffusion.

In the limit of small interior diffusivity (Fig.~S3a, default case), the dominant balance in the temperature budget is between vertical mean flow advection (red line) and eddy advection (cyan line), in which the horizontal components are dominant (not shown). The vertical diffusion term (dashed green line) only plays a minor role with sign changes in the vertical direction. This dominant balance is consistent with \citet{Wolfe-Cessi-2010:what}. When interior $\kappa_v$ is larger, the dominant balance becomes between vertical mean flow advection and explicit vertical diffusion, again consistent with \citet{Munk-1966:abyssal}. However, the two terms never perfectly balance each other even in the Kvx100 case where $\kappa_v$ is as unrealistically large as $10^{-3}$ m$^2$ s$^{-1}$, and their residual is balanced by eddy temperature transport. In the cases using large values of interior $\kappa_v$ (starting from Kvx20 case), the near-bottom temperature budget shows a dominant balance between horizontal mean flow advection and explicit vertical diffusion. A similar balance has been discussed by \citet{Mashayek-Ferrari-Nikurashin-Peltier-2015:influence} and is related to an abyssal layer where isopycnals intersect almost vertically with the bottom boundary.

In summary, we have so far shown that the isopycnal slopes in the Southern Ocean channel do respond to changes of vertical diffusivity only in the interior, even though the surface wind forcing is held constant in all the simulations. This means that SO eddies cannot set these slopes independently of other factors, and this conclusion is what we were after when determining our model experiment design. The exponential profile of $\sigma_2$ exists even when the interior $\kappa_v$ is very weak. However, the quality of the exponential fit of $\partial_z\sigma_2$ improves as the interior $\kappa_v$ is increased. And the stratification profiles ($\partial_z\sigma_2$) are not exponential in the low limit of interior $\kappa_v$.

\subsection{Southern Ocean Response}
\label{sec:SO}

\begin{figure}[t]
  \centering
  \noindent\includegraphics[width=0.75\paperwidth]{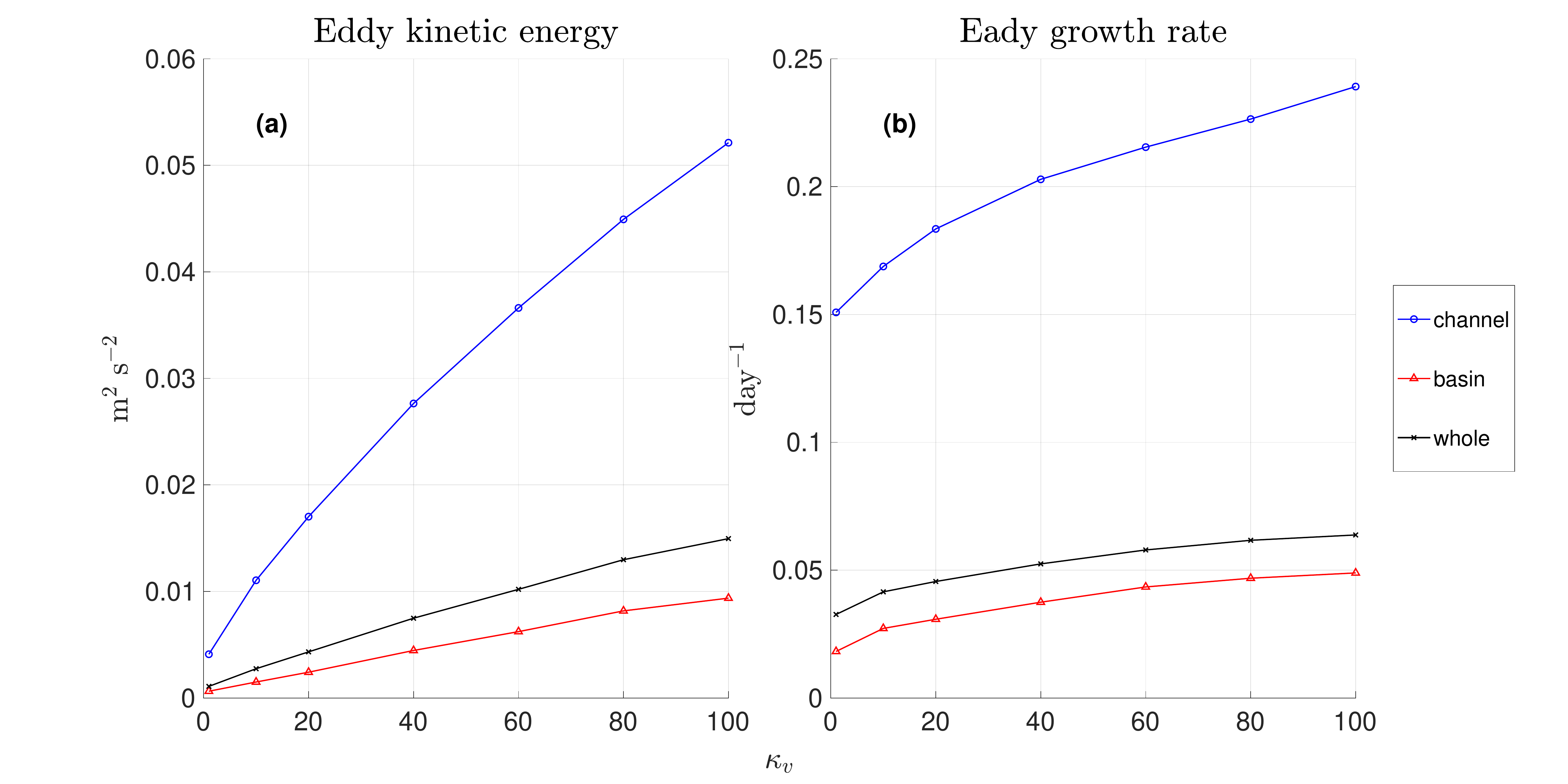}
  \caption{(a) Averaged eddy kinetic energy ($\sqrt{\frac{1}{2}\left( u^{\prime 2}+v^{\prime 2} \right)}$), as a function of interior $\kappa_v$ ($\times$ 10$^5$). (b) Averaged Eady growth rate ($0.31 f\left| \partial_z \mathbf{u}_H \right|/N$). Blue line with circles: in the channel; Red line with triangles: in the basin; black line with crosses: in the whole basin. }
  \label{EKE_EGR_line_plot}
\end{figure}

We have pointed out that the isopycnal slopes in the Southern Ocean channel become steeper as interior $\kappa_v$ is increased. The slopes of isopycnals determine the stability of the mean flow and steeper isopycnal slopes imply stronger instability and therefore more vigorous eddy activity in the channel. Eddy kinetic energy (EKE, $\sqrt{(u^{\prime 2}+v^{\prime 2})/2}$) averaged in different parts of the domain is shown in Fig.~\ref{EKE_EGR_line_plot}a. It is clear that eddy kinetic energy is increasing in the channel nearly linearly with interior $\kappa_v$, with smaller increase in the interior basin (away from the SO) as well. A similar phenomenon has been observed and explained in \citet{Munday-Johnson-Marshall-2013:eddy} that eddies become stronger when $\kappa_v$ increases because of steeper isopycnals in the channel which are a result of deeper isopycnals diffused farther downward in the interior. We note that \citet{Munday-Johnson-Marshall-2013:eddy} increase $\kappa_v$ uniformly over the domain (including the SO channel) and explore a smaller range of $\kappa_v$, but their explanation holds here as well, and this issue is briefly further explored next.

To explain what drives the increase of EKE in the Southern Ocean channel, we first use the Eady growth rate, defined as $0.31f\left| \partial_z \mathbf{u}_H \right|/N$ \citep{Vallis-2017:atmospheric} as a measure of baroclinic instability. Here, $f=2\Omega\sin\left( \theta \right)$ is the Coriolis parameter and $N^2=-g\partial_z\rho/\rho_0$ is the buoyancy frequency. Fig.~\ref{EKE_EGR_line_plot}b shows volume averaged Eady growth rate (per day) and it is clear that Eady growth rate increases with $\kappa_v$, and the increase is faster in the channel, similar to EKE (Fig.~\ref{EKE_EGR_line_plot}a).

Another way to analyze the increase of EKE is via the Lorenz energy cycle \citep[equations in appendix A, following][]{Storch-2012:estimate}. We find that the dominant terms in the Lorenz energy cycle in the Southern Ocean channel are the conversion from eddy to mean APE, $C(Pe,Pm)$, and from eddy APE to EKE, $C(Pe,Ke)$, shown in Fig.~\ref{LEC} ($C(A,B)$ is positive when $A$ is converting to $B$). Generally, mean available potential energy ($P_m$) is converted to eddy available potential energy ($P_e$), which is then converted to eddy kinetic energy ($K_e$) in the Southern Ocean channel in all simulations. The strength of the Lorenz energy cycle grows as the interior $\kappa_v$ is increased, consistent with the increase of EKE observed above. Fig.~S4 and S5 further show that, as $\kappa_v$ increases, more mean APE is injected into the Southern Ocean channel via surface buoyancy forcing ($G(P_m)$).

As a final approach to analyzing the change to the SO eddies, we consider a \cite{Gent-Mcwilliams-1990:isopycnal} type parameterization of the eddy-driven meridional circulation, where the eddy-driven stream function is assumed related to the isopycnal slopes $S_\rho$ as $\psi^*=\kappa_{GM}S_\rho$. As the SO eddies get stronger with interior $\kappa_v$, the diagnosed $\kappa_{GM}$ is also increasing (Appendix C, Fig.~S8), and shows a linear relation to interior $\kappa_v$ and averaged isopycnal slope in the Southern channel (except for the default case, Fig.~S8b,c). This implies that it may not be appropriate to assume a constant $\kappa_{GM}$ in simple box models for stratification and overturning circulation, and also the low-resolution ocean simulations.

To summarize, we find that the eddies in the Southern Ocean are responding to the interior $\kappa_v$ change, even though the wind forcing is unchanged. The eddy kinetic energy in the channel increases with interior $\kappa_v$ due to stronger baroclinic instability related to steeper isopycnal surfaces in the channel, and also due to a more vigorous Lorenz energy cycle that converts mean available potential energy to eddy available potential energy, and then to eddy kinetic energy.

\section{Conclusions}
\label{sec:conclusion}

Observations show that both the mean density profile and the mean stratification ($N^2$) profiles are very nearly exponential over large regions of the world ocean \citep[e.g.,][]{Miller-Yang-Tziperman-2019:reconciling}. A vertical advective-diffusive balance was proposed to explain the exponential profile of temperature in the Pacific Ocean by \citet{Munk-1966:abyssal}. This explanation assumed that deep water formation returns to the surface via diapycnal-mixing driven upwelling in the ocean interior. However, the required vertical diffusivity in the mid-depth ocean required in this scenario is an order of magnitude larger than the values revealed by tracer release experiments \citep{Ledwell-Watson-Law-1993:evidence, Polzin-Toole-Ledwell-Scmitt-1997:spatial}. This leaves the mid-depth exponential profile unexplained, and this is the focus of the present work.

Mixing observations also found non-uniform vertical mixing, concentrated near bottom and side ocean boundaries, due to tidal forcing and breaking of internal waves near topography \citep{Wunsch-Ferrari-2004:vertical}. Due to the boundary and bottom enhancement of vertical mixing, one might expect the diapycnal upwelling to be not uniform as assumed by \citet{Munk-1966:abyssal}, but concentrated where vertical mixing is enhanced, usually near ocean margins \citep[e.g,][]{Ferrari-Mashayek-McDougall-et-al-2016:turning}. The consequences of enhanced ocean margin mixing for the meridional overturning circulation were studied by \citet{Samelson-1998:large} and \citet{Scott-Marotzke-2002:location}. More relevant to this work, the consequences for what might be setting the mid-depth stratification were examined in \citet{Miller-Yang-Tziperman-2019:reconciling}. They showed that a Munk-like balance can hold in narrow boundary regions near ocean margins, with much larger vertical diffusion and vertical velocity there, resulting in a Munk-like stratification that is then advected toward the ocean interior where the diffusivity may be small, consistent with tracer release experiments and with a similar suggestion made by \citet{Munk-Wunsch-1998:abyssal}. The effect of bottom-intensified vertical mixing has been explored by \citet{Mashayek-Ferrari-Nikurashin-Peltier-2015:influence}. They show that a bottom layer has a different temperature budget: vertical diffusion is dominantly balanced by  horizontal advection.

An alternative to the above diffusive ideas is often referred to as the ``adiabatic MOC''. According to this scenario, deep water formation is balanced by along-isopycnal upwelling due to wind forcing in the Southern Ocean, rather than by diapycnal fluxes in the ocean interior \citep{Marshall-Speer-2012:closure}, although a recent paper used ECCO data to point out that abyssal diapycnal transport dominates the adiabatic transport in the SO \citep{Rousselet-Cessi-Forget-2021:coupling}. The water makes it from deep water formation sites all the way to the SO along isopycnals. In this picture, Southern Ocean eddies were proposed to play a dominant role in determining the interior stratification \citep{Wolfe-Cessi-2010:what, Nikurashin-Vallis-2011:theory, Nikurashin-Vallis-2012:theory}. The eddies control the isopycnal slopes in the SO, and the sloping SO isopycnals then connect the surface density distribution of the Southern Ocean to a vertical distribution in the closed basin, thus setting the mid-depth interior stratification. Model studies show that with this mechanism included, the interior stratification does not vanish in the absence of vertical mixing \citep{Wolfe-Cessi-2010:what}, in contrast to the prediction of the Munk-like ideas. The difficulty with this adiabatic MOC idea, though, is that the resulting vertical stratification within the ocean interior away from the SO is very far from exponential as shown in \citet{Miller-Yang-Tziperman-2019:reconciling} as well as in Figure~\ref{rho_z_exp_fit}a above.

In this paper, we used idealized configurations of MITgcm, at an eddy-permitting resolution to examine if SO eddies indeed determine the interior mid-depth stratification rather than adjust to it. We did so by varying vertical diffusivity only in the basin interior away from the SO. The eddies are expected to control the SO slopes by balancing the wind via either a near-vanishing residual circulation, or by bringing the slopes to marginal criticality and both options are further discussed below. We found, however, that both SO isopycnal slopes and the interior stratification respond to interior vertical diffusivity even as the wind forcing and Southern Ocean vertical mixing are unchanged. As interior vertical mixing increases, the SO slopes steepen and vertical profiles of stratification ($\partial_z \sigma_2$) becomes more exponential. These large diffusivity cases considered here may be physically represented in the ocean by a scenario of high vertical diffusion near ocean margins only \citep{Miller-Yang-Tziperman-2019:reconciling}. We conclude that the SO isopycnal slopes are not uniquely determined by the eddies, but rather, that eddies can adjust to the slopes that are determined externally by the combination of the SO surface and ocean interior density distributions. As the vertical diffusivity is increased, the dominant balance in the temperature budget away from the SO changes from one between vertical mean flow advection and eddy advection \citep{Wolfe-Cessi-2010:what}, to one similar to the advective-diffusive balance \citep{Munk-1966:abyssal}, with a minor but non-negligible role played by eddy transport.

Consider in some more depth the two possible mechanisms by which SO eddies can determine the isopycnal slopes there. First, the northward Ekman transport driven by westerlies over the Southern Ocean can only return poleward below the sill depth where a geostrophically-balanced zonal pressure gradient may develop. This wind-induced overturning steepens isopycnal surfaces, which produces eddies. These eddies act to flatten the isopycnal surfaces by drawing APE from the slopes, and this competition between wind forcing and eddies will determine the slope of the isopycnal surfaces which will map the SO surface density distribution to the interior. When the wind forcing exactly balances the effect of the eddies, the residual circulation vanishes, leading to a condition on the slope magnitudes \citep[e.g.,][]{Marshall-Radko-2006:model}. We calculated the residual circulation for the different interior diffusivity runs and found that it varies and does not necessarily vanish. At the same time, the values of effective $\kappa_{GM}$ also increases as eddies get stronger in the channel with increasing interior diffusivity. In the simple box models of stratification, a constant Gent-McWilliams type horizontal mixing coefficient is usually assumed \citep[e.g.,][]{Gnanadesikan-1999:simple,Nikurashin-Vallis-2011:theory,Jones-Cessi-2016:interbasin}. The response we see of SO eddy activity to interior vertical diffusivity suggests limits to this simple assumption.

An alternative way to examine the role of eddies in setting SO isopycnal slopes involves the issue of marginal criticality. In this picture, as the external processes generating the slopes get stronger, the SO isopycnal slopes steepen. Stronger baroclinic instability make eddies become stronger, draw APE from the slopes and act to flatten them, until the slopes are back to marginal criticality \citep{Stone-1978:baroclinic,Jansen-Ferrari-2012:macroturbulent,Jansen-Ferrari-2013:equilibration,Jansen-Ferrari-2013:vertical}. At this point the processes generating the slopes exactly balance the effect of the eddies. However, in our runs, Southern Ocean eddies do not maintain a mean state of marginal criticality \citep[as quantified the ratio between the Rhines scale and the deformation radius, following][]{Jansen-Ferrari-2013:equilibration}.

We conclude that Southern Ocean isopycnal slopes are not determined only by eddies as the theories of marginal criticality or vanishing residual overturning circulation suggest. Instead, SO eddies, and therefore also the isopycnal slopes there, respond to changes in interior stratification as well. This means that the communication between the Southern Ocean and the closed basin is not one-way (eddies to mid-depth stratification), and that changes in the interior stratification can also trigger responses in the Southern Ocean.

There are several caveats to note. First, the model configuration is an idealization of the Atlantic Ocean (surface buoyancy boundary conditions induces deep water formation at the northern high latitudes), but the advective-diffusive balance is expected to work better in the Pacific Ocean where the upper cell of the overturning circulation does not exist. Second, we only use a single-basin configuration in this work, the communication between the Atlantic-type basin (with deep water formation) and Pacific-type basin (without deep water formation) is missing. Third, it is artificial to only change the value of vertical mixing in the interior, of course, although this strategy was adopted here to successfully answer a very specific question rather than used for a simulation of a realistic parameter regime. Some of the diffusivity values we used in this study are also unrealistically large. However, again, this design is consistent with the purpose of this study to find out whether the Southern Ocean is responsive to interior stratification changes. 

\section*{Appendix A: Lorenz Energy Cycle}
Following \citet{Storch-2012:estimate}, we calculate different terms in the Lorenz energy cycle using the following equations.

For mean kinetic energy and eddy kinetic energy,
\begin{equation}\label{LEC_kinetifc}
  \begin{split}
    &K_m=\iiint_V\frac{1}{2}\rho_0 (\overline{u}^2+\overline{v}^2)dv,\\
    &K_e=\iiint_V\frac{1}{2}\rho_0\overline{\left( u^{\prime 2}+v^{\prime 2} \right)}dv,
  \end{split}
\end{equation}
in which $u$ and $v$ are eastward and northward velocities respectively, and $rho_0$  is a constant density which is 1025 kg m$^{-3}$ in our calculation.

And mean available potential energy and eddy available potential energy are defined as
\begin{equation}\label{LEC_potential}
  \begin{split}
    &P_m=-\iiint_v\frac{g}{2\partial_z \rho_{\text{ref}}}\overline{\rho^*}^2dv\\
    &P_e=-\iiint_v\frac{g}{2\partial_z \rho_{\text{ref}}}\overline{\rho^{*\prime 2}}dv,
  \end{split}
\end{equation}
in which $g=9.8$ m s$^{-2}$ is gravitational acceleration. $\rho_{\text{ref}}$ is a reference density profile which is taken to be the horizontal average between 30$^\circ$S to 30$^\circ$N and three degrees from the western and eastern boundaries in our calculation. $\rho^*=\rho-\rho_{\text{ref}}$ is the departure of density from the reference density profile.

The time evolution of the energy terms can be written as
\begin{equation}\label{LEC}
  \begin{split}
    &\frac{dP_m}{dt}=C(P_e,P_m)-C(P_m,K_m)+G(P_m)-D(P_m)\\
    &\frac{dP_e}{dt}=-C(P_e,P_m)-C(P_e,K_e)+G(P_e)-D(P_e)\\
    &\frac{dK_m}{dt}=C(K_e,K_m)+C(P_m,K_m)+G(K_m)-D(K_m)\\
    &\frac{dK_e}{dt}=-C(K_e,K_m)+C(P_e,K_e)+G(K_e)-D(K_e),
  \end{split}
\end{equation}
in which the $C$ terms are conversion between different energy terms, $G$ is generation rate due to surface forcing, and $D$ is dissipation term. The $D$ terms are usually calculated as a residual to close the energy budgets.

As in eq.~A3 the conversion between energy terms are defined below ($C(A,B)$ is positive if $A$ is converted to $B$)
\begin{equation}\label{LEC_convert}
  \begin{split}
    &C(P_e,P_m)=-\iiint_V\frac{g}{\partial_z\rho_{\text{ref}}}\overline{\rho^\prime\mathbf{u}_h^\prime}\cdot\nabla_h\overline{\rho}dv\\
    &C(K_e,K_m)=\iiint_v\rho_0\left( \overline{u^\prime\mathbf{u}_H^\prime}\cdot\nabla\overline{u}+\overline{v^\prime\mathbf{u}_H^\prime}\cdot\nabla\overline{v}  \right)dv\\
    &C(P_m,K_m)=-\iiint_vg\overline{\rho}\overline{w}dv\\
    &C(P_e,K_e)=-\iiint_vg\overline{\rho^\prime w^\prime}dv.
  \end{split}
\end{equation}

Finally, the energy generation terms in eq.~A3 are defined as
\begin{equation}
  \begin{split}
    &G(P_m)=-\iint_Sg\left( \frac{\alpha_s}{C_{p,s}\partial_z\rho_{\text{ref},s}}\overline{Q_s}\cdot\overline{\rho^*_s}+\frac{\beta_s}{\partial_z\rho_{\text{ref,s}}}\overline{F_{\text{salt}}}\cdot\overline{\rho^*} \right)dxdy\\
    &G(P_e)=-\iint_Sg\left( \frac{\alpha_s}{C_{p,s}\partial_z\rho_{\text{ref},s}}\overline{Q_T^\prime\rho^{*\prime}_s}+\frac{\beta_s}{\partial_z\rho_{\text{ref,s}}}\overline{F^\prime_{\text{salt}}\rho^{*\prime}} \right)dxdy\\
    & G(K_m)=\iint_S\overline{\mathbf{\tau}}\cdot\overline{\mathbf{u}_{H,s}}dxdy\\
    & G(K_e)=\iint_S\overline{\mathbf{\tau}^\prime\cdot\mathbf{u}^\prime_{H,s}}dxdy
  \end{split}
\end{equation}
in which subscript $s$ means evaluation at the surface layer. $\alpha$ and $\beta$ are thermal expansion and haline contraction coefficients. $C_p$ is the isobaric heat capacity, $Q_T$ is surface heat flux and $F_{\text{salt}}$ is surface salt flux. $\mathbf{\tau}$ stands for surface wind forcing.

\section*{Appendix B: Length scales in the Southern Ocean channel and criticality}

Following \citet{Jansen-Ferrari-2012:macroturbulent}, we define the deformation radius as
\begin{equation}
  R_d=2\int_{\mathrm{bot}}^{\mathrm{top}}\frac{NdZ}{f},
\end{equation}
in which $f=2\omega\sin(\theta)$ is the Coriolis parameter, and $N^2=-g\partial_z\rho/\rho_0$ is the buoyancy frequency.

And we define the Rhines scale at which the upper-scale energy cascade should halt as
\begin{equation}
  R_b=2\pi\frac{\mathrm{EKE}_t^{1/4}}{\beta^{1/2}},
\end{equation}
in which $\mathrm{EKE}_t$  is the barotropic eddy kinetic energy, and $\beta=2\omega\cos(\theta)/R$ is the meridional gradient of the Coriolis parameter. The criticality parameter is further defined as the ratio between $R_b$ and $R_d$, averaged in the SO channel.

\section*{Appendix C: Residual overturning circulation and horizontal eddy diffusivity}

The residual overturning circulation on a surface of constant buoyancy $b_0$ is defined as
\begin{equation}
  \tilde{\psi}(y,b_0)=-\frac{1}{T}\int_{0}^T\int_{0}^{L_x}\int_{\mathrm{bot}}^{\mathrm{top}}v(x,y,z,t)\mathcal{H}\left( b(x,y,z,t)-b_0 \right)dzdxdt,
\end{equation}
in which $T$ is time, $L_x$ is the zonal extent of the basin, $v$ is meridional velocity and $\mathcal{H}$ is the Heaviside function.

And the position of the surface of constant buoyancy $b_0$ is further determined as
\begin{equation}
  Z(y,b_0)=\frac{1}{T}\int_{0}^T\int_{0}^{L_x}\int_{\mathrm{bot}}^{\mathrm{top}}\mathcal{H}\left( b(x,y,z,t)-b_0 \right)dzdxdt.
\end{equation}

The residual overturning circulation can then be interpolated to depth coordinate using the $Z$ field (Fig.~S7).

The residual circulation in the channel can be decomposed into a wind-induced part and an eddy-induced part
\begin{equation}
  \tilde{\psi}=\overline{\psi}+\psi^*=\frac{\tau}{\rho_0\left| f \right|}+
  \frac{\overline{v^\prime b^\prime}}{\partial_z \overline{b}}\approx \frac{\tau}{\rho_0\left| f \right|}-\kappa_{GM}\frac{\partial_y\overline{b}}{\partial_z\overline{b}}=\approx\frac{\tau}{\rho_0\left| f \right|}+\kappa_{GM}S_\rho,
\end{equation}
with a Gent-McWilliams-type parameterization assumed. We calculate the meridional profiles of $\kappa_{GM}$ using the averaged residual overturning circulation and isopycnal slope in the channel (Fig.~S8).

%%%%%%%%%%%%%%%%%%%%%%%%%%%%%%%%%%%%
% TABLES
%%%%%%%%%%%%%%%%%%%%%%%%%%%%%%%%%%%%
\newpage\clearpage
\begin{table}[t]
\caption{Summary of the numerical simulations.}\label{case_series}
\begin{center}
\begin{tabular}{ccc}
\hline\hline
Case No. & Code name & Interior $\kappa_v$ \\
\hline
 1 & control & 10$^{-5}$  \\
 2 & Kvx10 & 10$^{-4}$  \\
 3 & Kvx20 & 2$\times$ 10$^{-4}$  \\
 4 & Kvx40 & 4$\times$ 10$^{-4}$  \\
 5 & Kvx60 & 6$\times$ 10$^{-4}$  \\
 6 & Kvx80 & 8$\times$ 10$^{-4}$  \\
 7 & Kvx100 & 10$^{-3}$  \\
\hline\hline
\end{tabular}
\end{center}
\end{table}

\begin{table}[t]
\caption{Other numerical parameters}\label{model_param}
\begin{center}
\begin{tabular}{ccccc}
\hline
  $A_h$ & $A_4$ & $A_r$ & $\kappa_4$ & $r_b$ \\
\hline
  10 m$^2$ s$^{-1}$  & 1.5$\times$ 10$^{10}$ m$^4$ s$^{-1}$ & 10$^{-4}$ m$^2$ s${-2}$ & 1.5$\times$ 10$^{10}$ m$^4$ s$^{-1}$ & 2$\times$ 10$^{-4}$ s$^{-1}$ \\
\hline
\end{tabular}
\end{center}
\end{table}

%%%%%%%%%%%%%%%%%%%%%%%%%%%%%%%
% FIGURES
%%%%%%%%%%%%%%%%%%%%%%%%%%%%%%%
\newpage\clearpage

\newpage\clearpage

\newpage\clearpage

\newpage\clearpage

\newpage\clearpage

\newpage\clearpage

\newpage\clearpage
\begin{figure}[t]
  \centering
  \noindent\includegraphics[width=0.75\paperwidth]{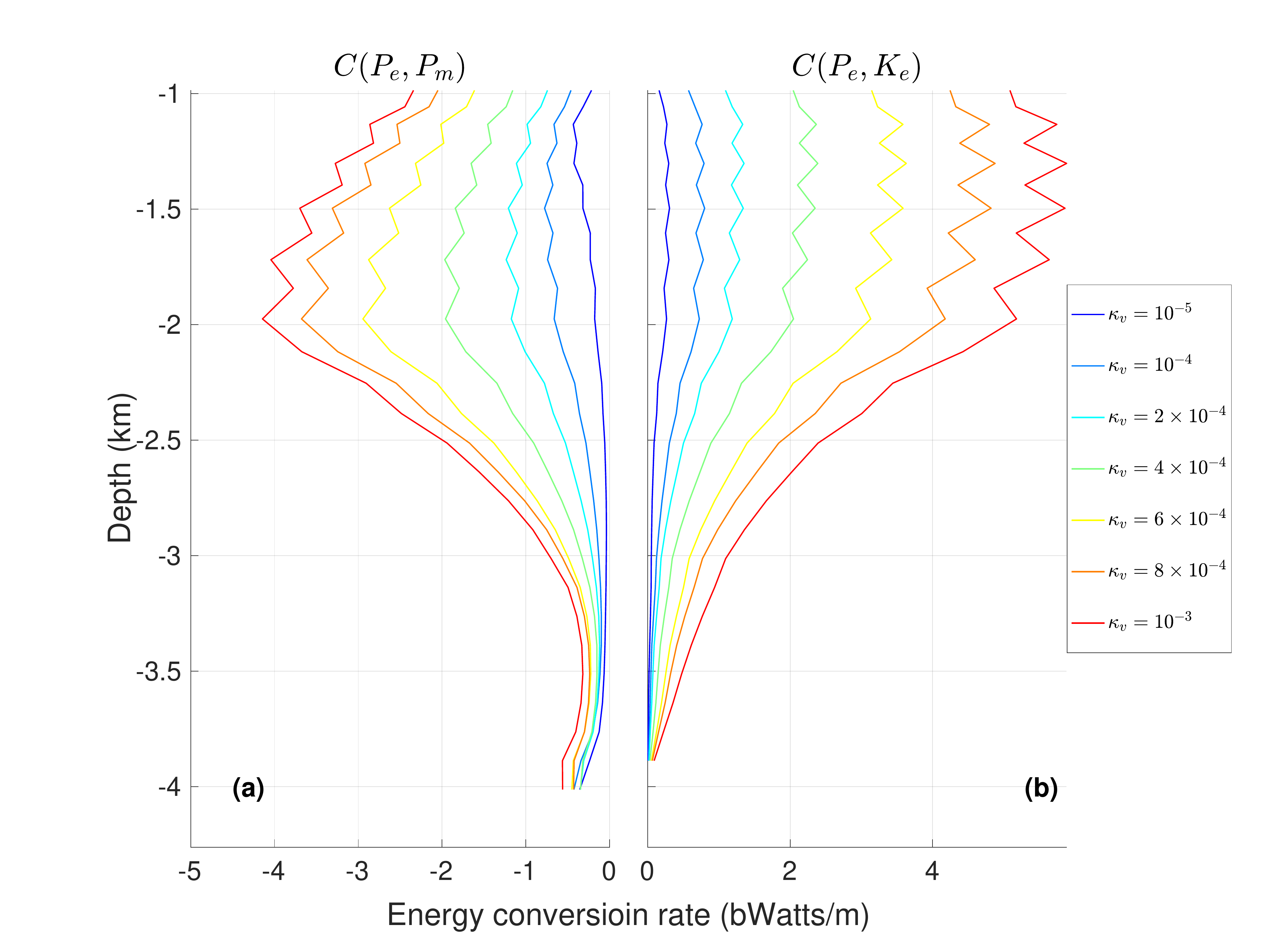}
  \caption{The vertical profiles of the two dominant terms in the Lorenz energy cycle in the channel of each case. (a) $C(P_e,P_m)$: conversion from eddy available potential energy(APE) to mean APE; (b) $C(P_e, K_e)$: conversion from eddy APE to eddy kinetic energy.}
  \label{LEC}
\end{figure}

\newpage\clearpage
\setcounter{figure}{0}
\renewcommand{\thefigure}{S\arabic{figure}}

\section*{Supplementary Figures}

\begin{figure}[h]
  \centering
  \noindent\includegraphics[width=0.75\paperwidth]{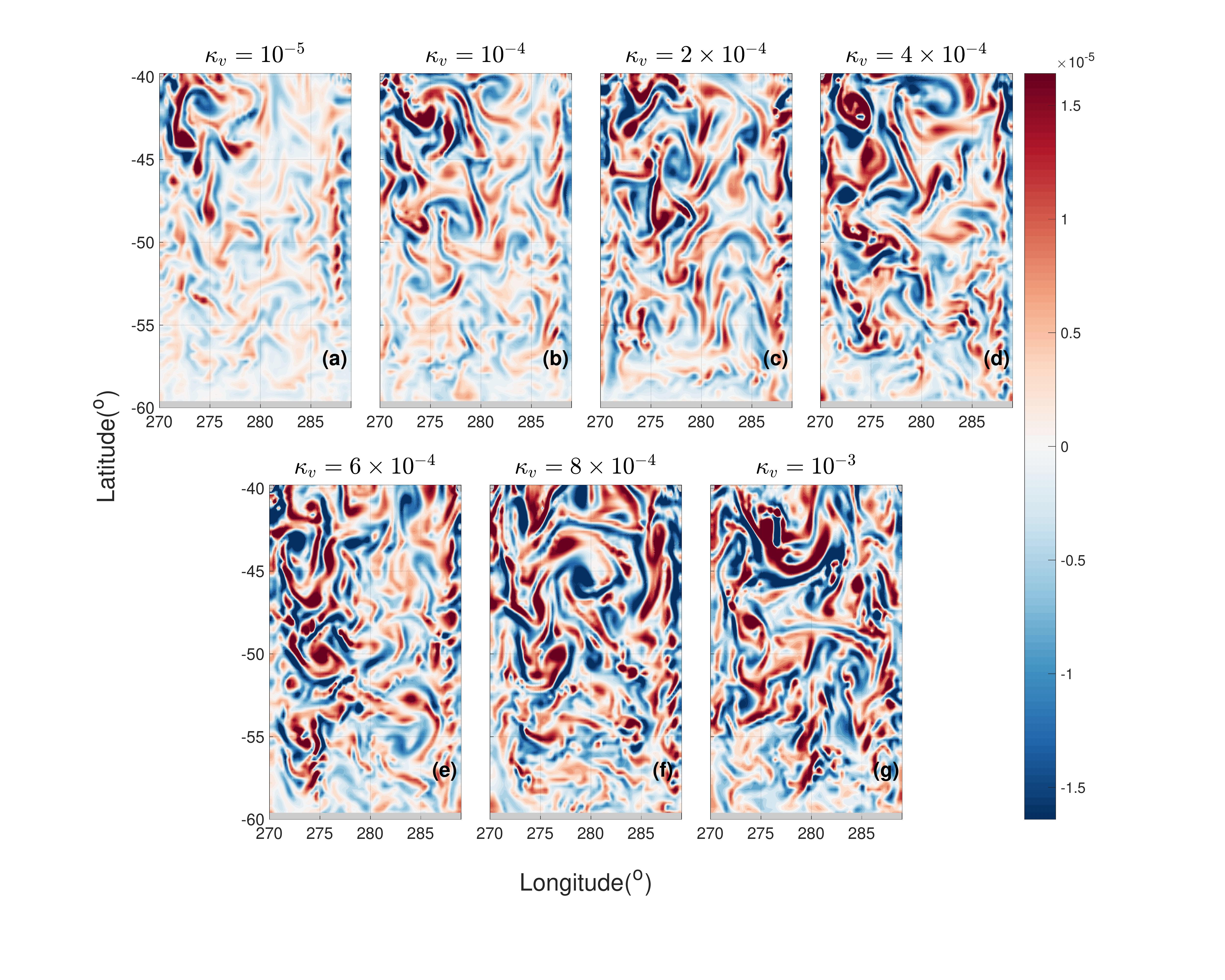}
  \caption{A snapshot of surface relative vorticity in the southern channel. (a) control; (b) Kvx10; (c) Kvx20; (d) Kvx40; (e) Kvx60; (f) Kvx80; (g) Kvx100.}
  \label{snapshot}
\end{figure}

\newpage\clearpage
\begin{figure}[t]
  \centering
  \noindent\includegraphics[width=0.75\paperwidth]{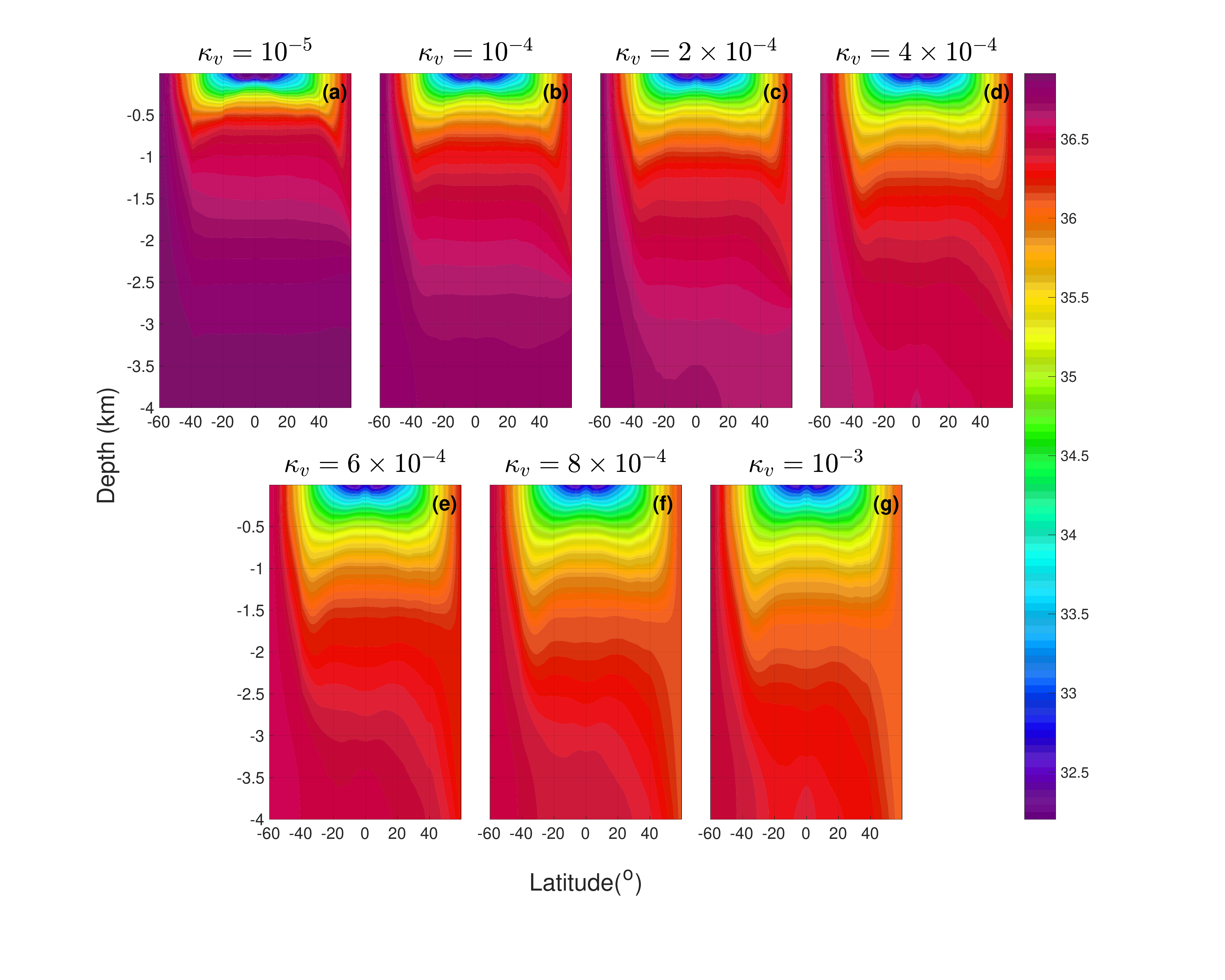}
  \caption{Zonally averaged $\sigma_2$ for the whole latitudinal range of the model domain for each simulation. (a) control; (b) Kvx10; (c) Kvx20; (d) Kvx40; (e) Kvx60; (f) Kvx80; (g) Kvx100.}
  \label{strat_full}
\end{figure}

\newpage\clearpage
\begin{figure}[t]
  \centering
  \noindent\includegraphics[width=0.75\paperwidth]{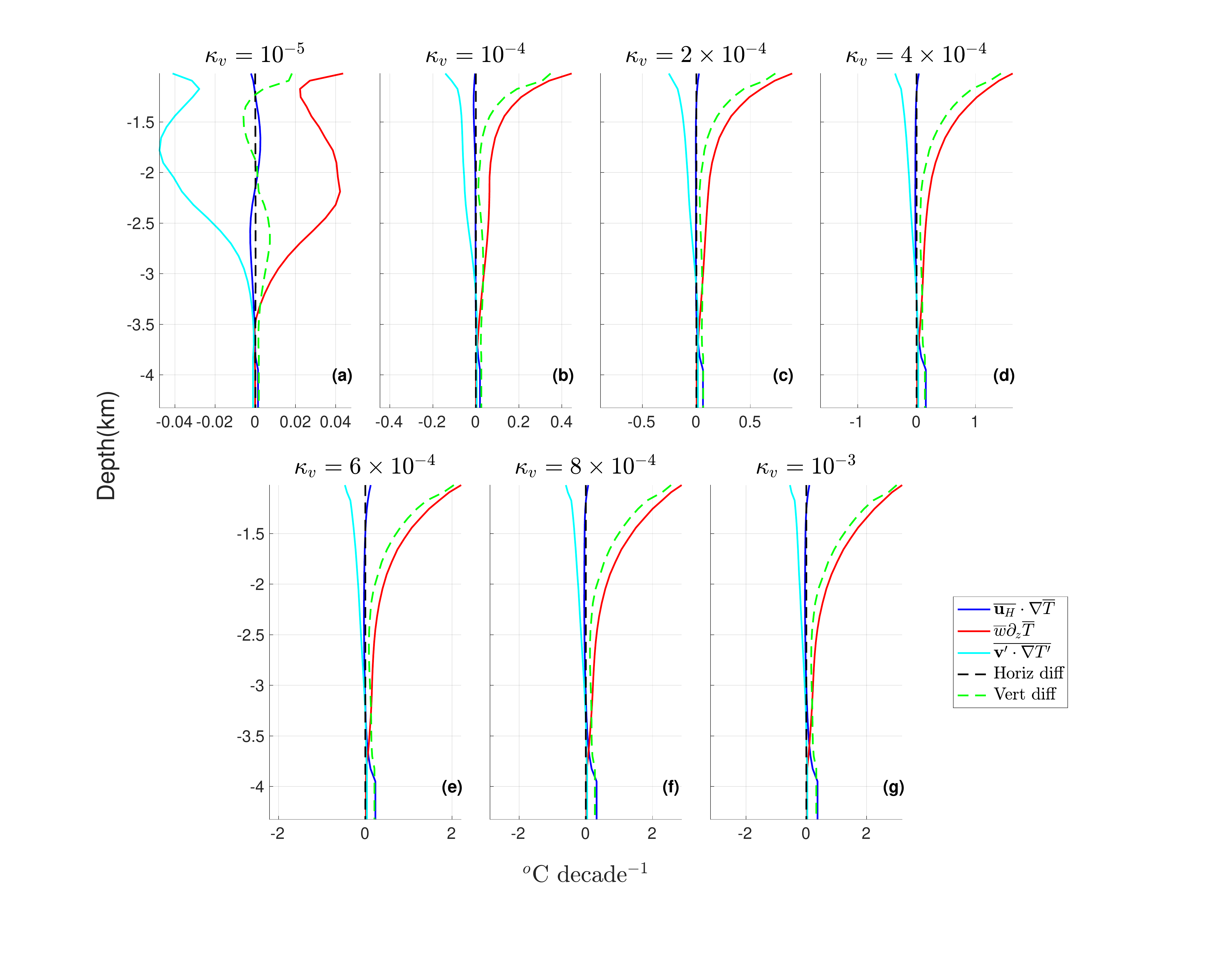}
  \caption{Vertical profiles for different terms in the temperature budget $\overline{\mathbf{u}_H}\cdot\nabla\overline{T}+\overline{w}\partial_z\overline{T}+\overline{\mathbf{v}^\prime\cdot\nabla T^\prime}=\text{Horiz diff}+\text{Vert diff}$. Solid lines are for terms on the left hand side and dashed lines are for terms on the right hand side. The vertical profiles are averaged from 30$^\circ$ S to 30$^\circ$ N, three degrees from the western and the eastern boundaries. Solid blue line: horizontal mean flow advection; solid red line: vertical mean flow advection; solid cyan line: eddy advection; dashed black line: explicit horizontal diffusion; dashed green line: explicit vertical diffusion.  (a) control; (b) Kvx10; (c) Kvx20; (d) Kvx40; (e) Kvx60; (f) Kvx80; (g) Kvx100. }
  \label{Temp_budget}
\end{figure}

\newpage\clearpage
\begin{figure}[t]
  \centering
  \noindent\includegraphics[width=0.75\paperwidth]{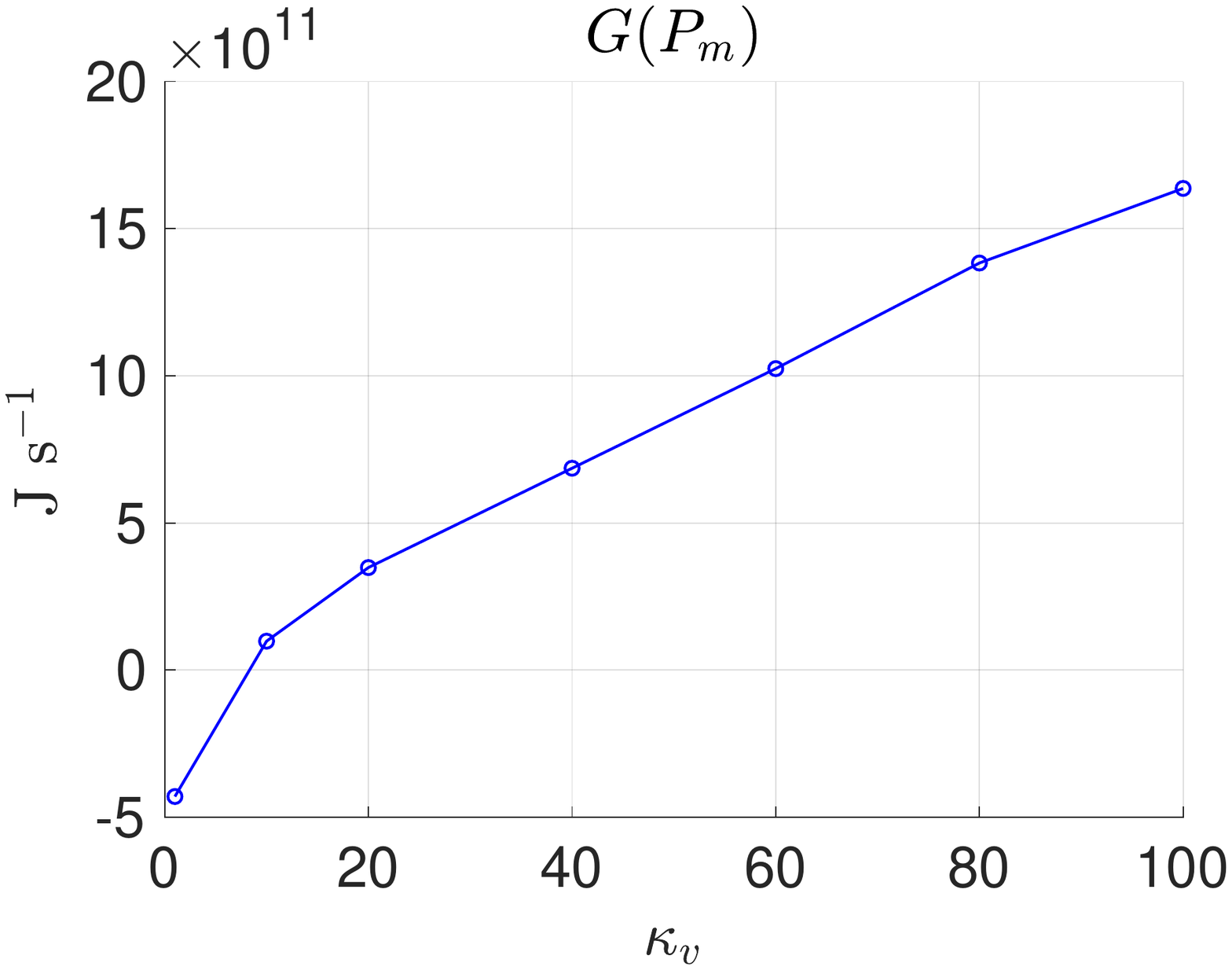}
  \caption{The generation rate of mean available potential energy ($G(Pm)$), integrated in the southern channel as a function of interior $\kappa_v$ ($\times 10^5$) used.}
  \label{LEC_generate}
\end{figure}

\newpage\clearpage
\begin{figure}[t]
  \centering
\noindent\includegraphics[width=0.75\paperwidth]{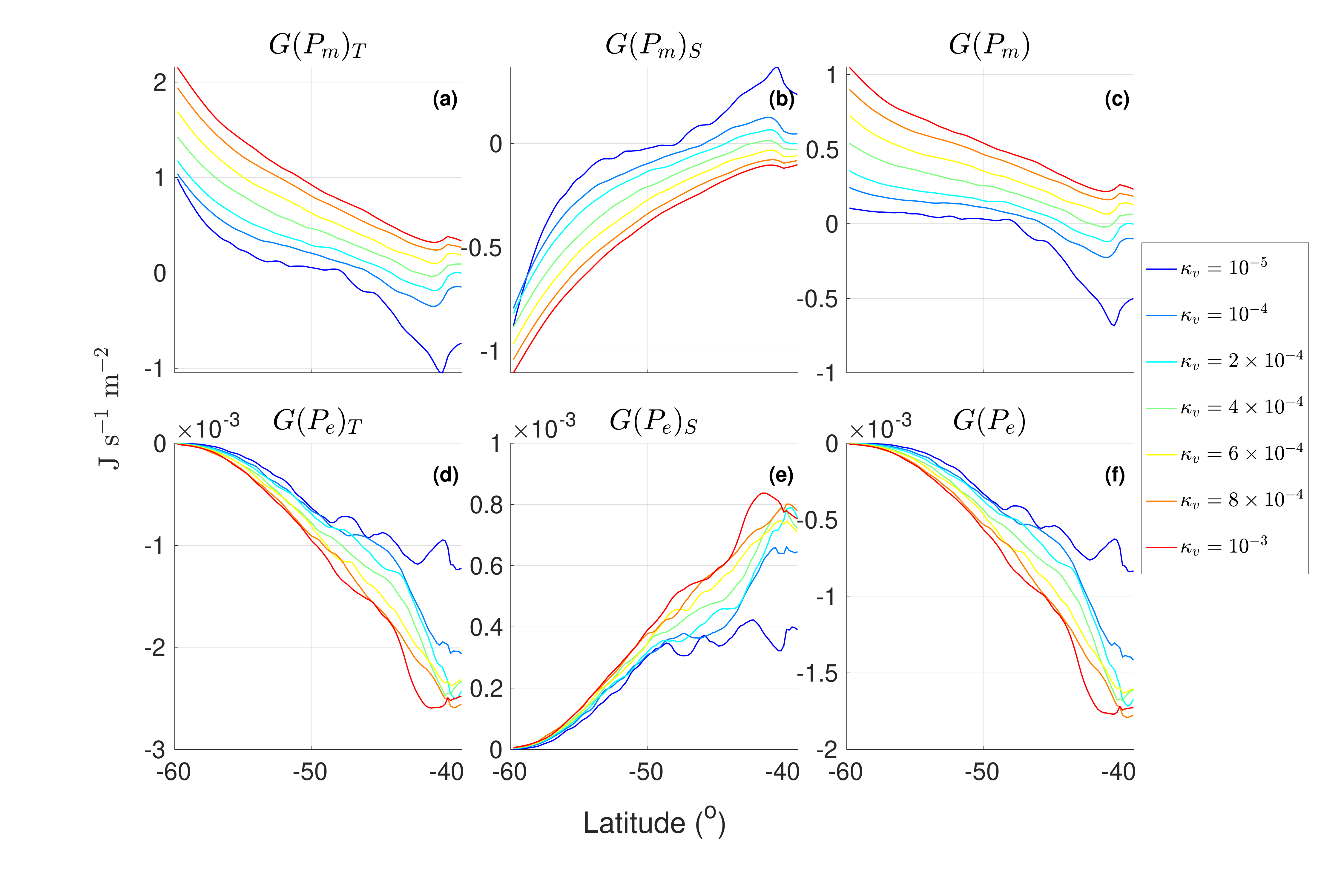}
  \caption{The generation rate of mean/eddy APE in the southern channel as function of latitude for all simulations (color). (a) Generation rate of mean APE induced by surface heat flux; (b) generation rate of mean APE induced by surface salinity flux; (c) total generation rate of mean APE; (d) generation rate of eddy APE induced by surface heat flux; (e) generation rate of eddy APE induced by surface salinity flux; (f) total generation rate of eddy APE.}
  \label{LEC_surf}
\end{figure}

\newpage\clearpage
\begin{figure}[t]
  \centering
\noindent\includegraphics[width=0.75\paperwidth]{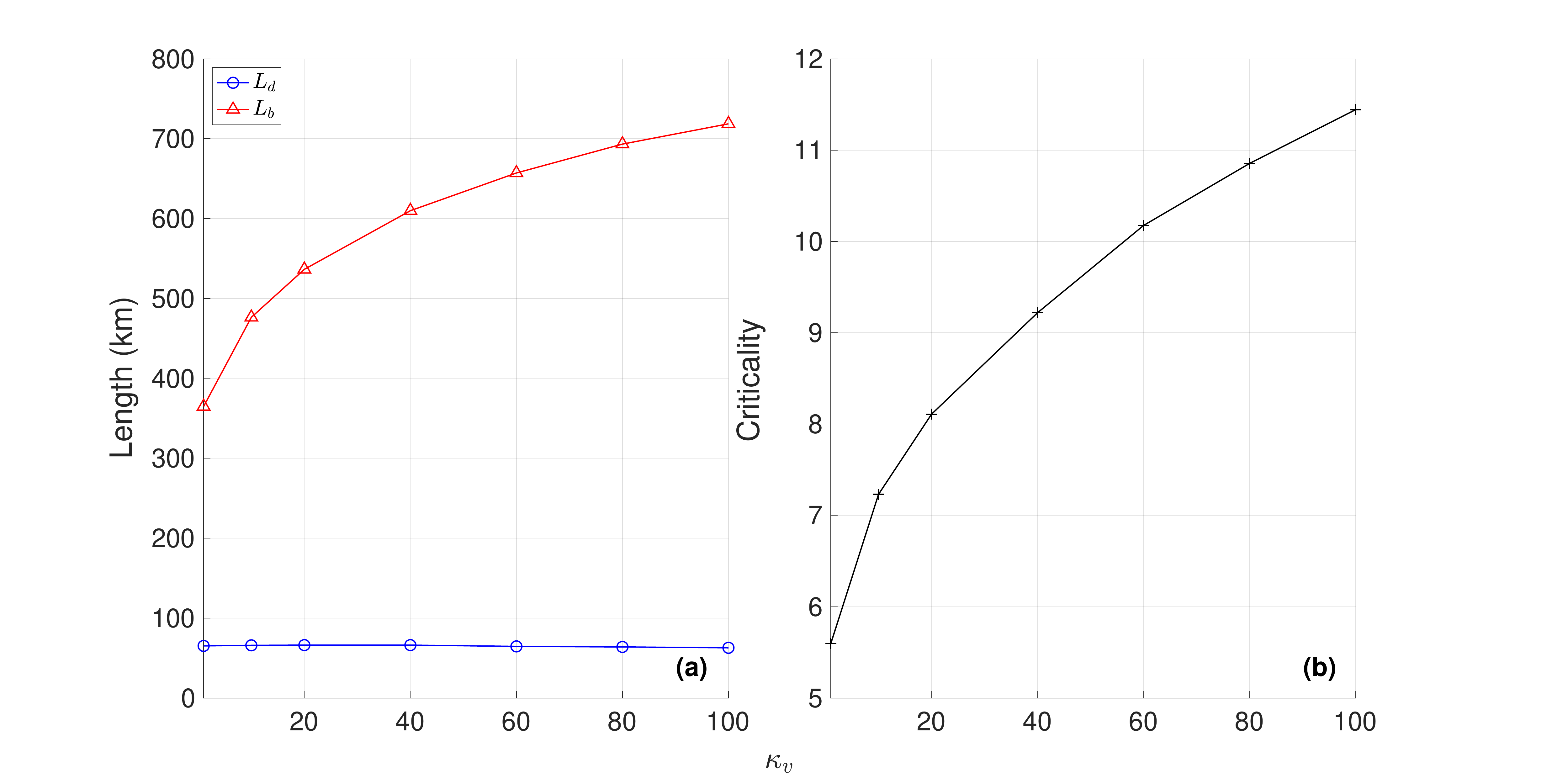}
  \caption{Measurement of Southern Ocean criticality. (a) Length scales in the Southern Ocean: (Blue dotted line) Rossby deformation radius; (red triangle line) Rhines scale; (b) criticality parameter: the ratio between Rhines scale and Rossby deformation radius.}
  \label{criticality}
\end{figure}

\newpage\clearpage
\begin{figure}[t]
  \centering
\noindent\includegraphics[width=0.75\paperwidth]{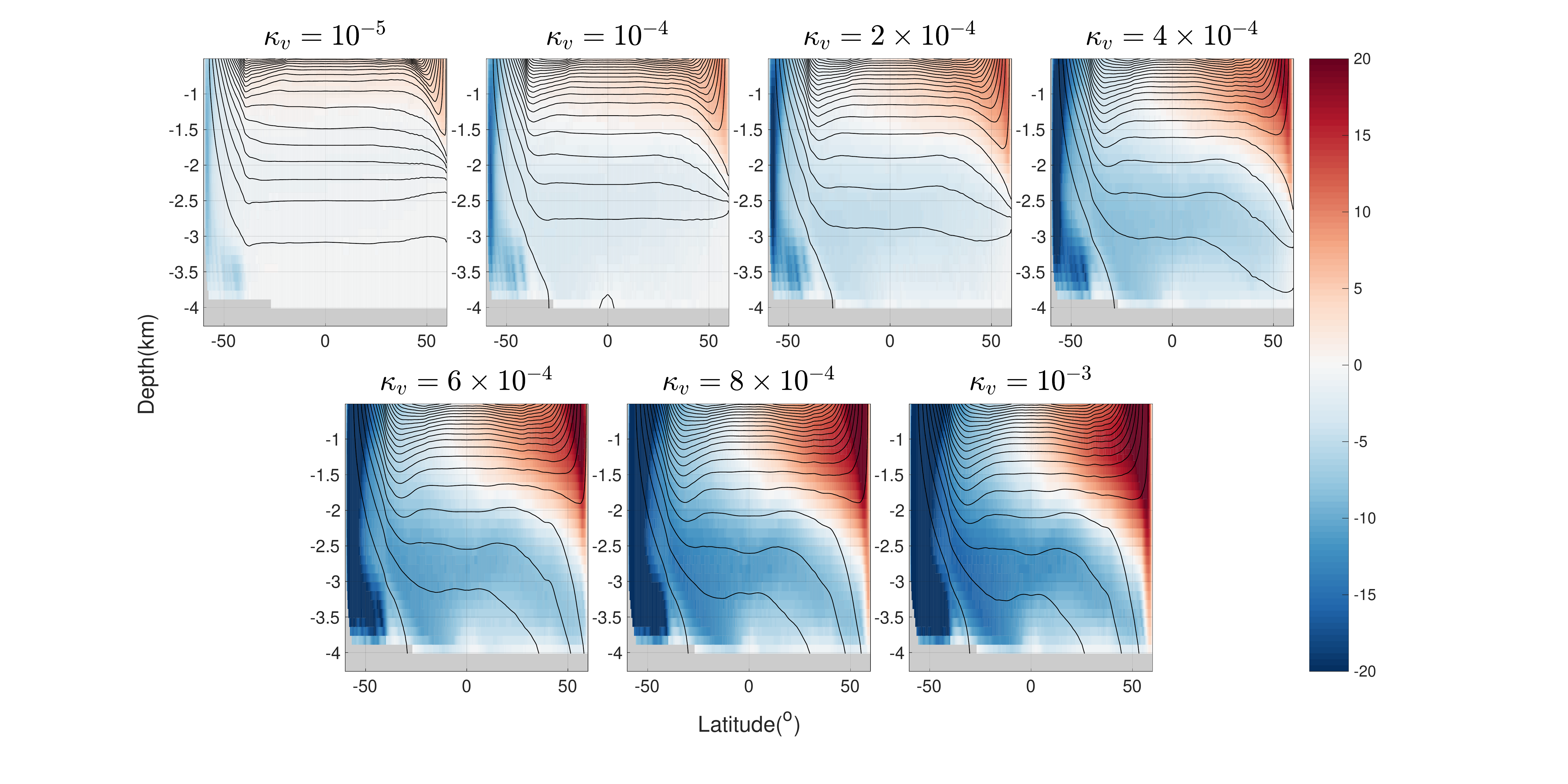}
  \caption{Residual meridional overturning circulation, for each simulation.}
  \label{MOC}
\end{figure}

\newpage\clearpage
\begin{figure}[t]
  \centering
\noindent\includegraphics[width=0.75\paperwidth]{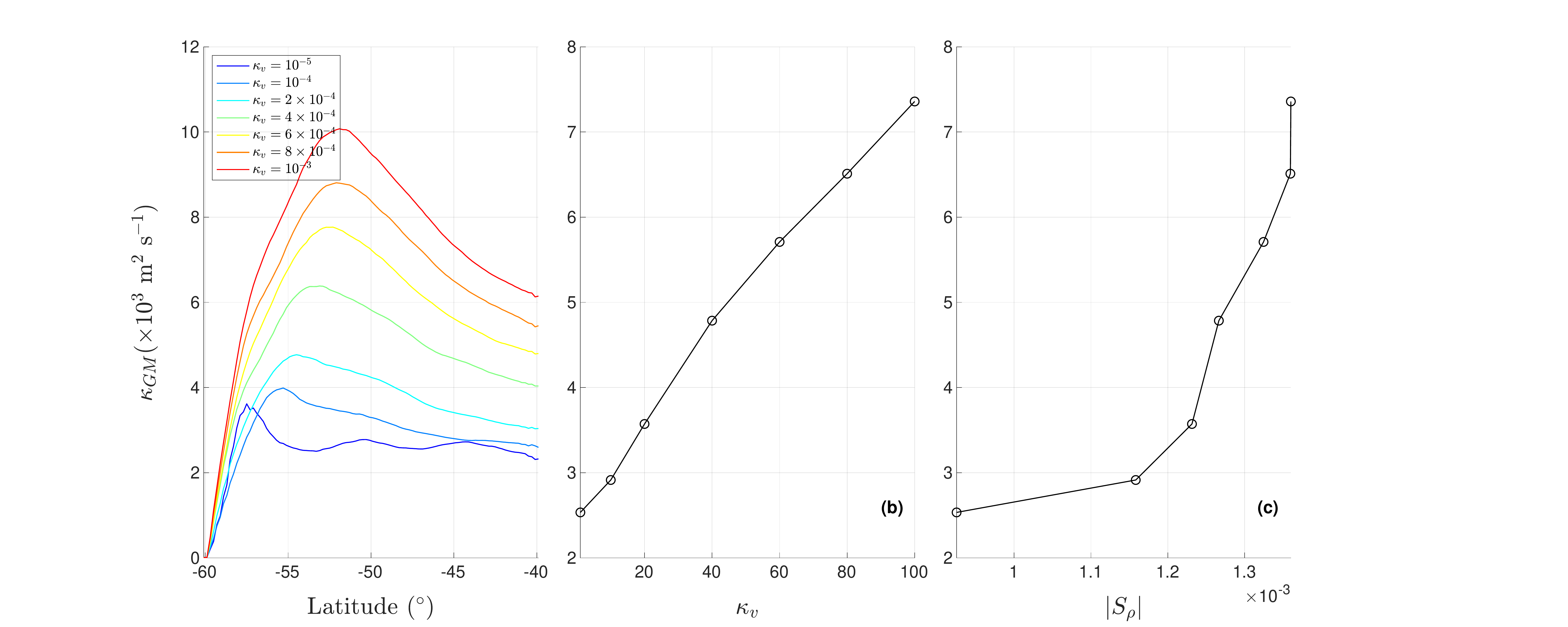}
  \caption{Gent-McWilliams diffusivity coefficient. (a) $\kappa_{GM}$ meridional profiles for each case; (b) averaged $\kappa_{GM}$ as function of interior $\kappa_v$; (c) averaged $\kappa_{GM}$ as function of the absolute value of the averaged isopycnal slope $S_{\rho}$.}
  \label{Kappa_GM}
\end{figure}

\newpage\clearpage
\bibliographystyle{apalike}
\bibliography{export}

\end{document}